\documentclass[12pt]{article}
\usepackage{jheppub}
\usepackage{amsmath,amssymb,bm}
\usepackage{gensymb}
\usepackage{nicefrac} 
\usepackage{epsfig}
\usepackage{graphicx}
\usepackage{slashed}
\usepackage{epsfig} \usepackage{graphicx} \usepackage{color}
\usepackage{mathrsfs} \usepackage{amssymb} \usepackage{amsmath} \usepackage{url}

\usepackage{xspace} 

\usepackage{braket}
\usepackage{hyperref}

\def \PPbr{$P^0-\bar{P}^0$ \ }

\def \oln{\bar}
\def \widebar{\bar}
\def \pl{\parallel}
\def \ls{{\lambda\sigma}}
\def \c2b{\cos 2\beta}
\def \s2b{\sin 2\beta}

\def \pp{\perp\perp}

\def\T {\ensuremath{T}\xspace}
\def\CP {\ensuremath{C\!P}\xspace}
\def\CPT {\ensuremath{C\!P T}\xspace}

\def\beq{\begin{equation}}
\def\eeq{\end{equation}}
\def\bea{\begin{eqnarray}}
\def\eea{\end{eqnarray}}
\def\nn{\nonumber}

\def\roughly#1{\mathrel{\raise.3ex\hbox
		{$#1$\kern-.75em\lower1ex\hbox{$\sim$}}}}

\def\Re{\text{Re}}
\def\Im{\text{Im}}

\allowdisplaybreaks
\title{
Dealing with T and CPT violations in mixing as well as direct and indirect CP violations for neutral mesons decaying to two vectors}

\author[a]{Anirban Karan.}
\affiliation[a]{Indian Institute of Technology Hyderabad, Kandi,  Sangareddy-502285, Telengana, India}

\emailAdd{kanirban@iith.ac.in}
\abstract{}
\begin{document}

\abstract{A large number of observables can be constructed from differential decay rate based on the polarization of final state while considering decay of a neutral meson $(P^0 \text{ or } \bar P^0)$ to two vector particles. But all of these observables are not independent to each other since there are only a few independent theoretical parameters controlling the whole dynamics and therefore various relations among observables emerge. In this paper, we have studied the behaviour of observables for neutral meson decaying to two vectors in presence of \T and \CPT violations in mixing accompanied by both direct and indirect \CP violations. We have expressed all of the fourteen unknown theoretical parameters for this scenario in terms observables only and constructed the complete set of thirty four relations among observables whose violation would signify the existence of some new Physics involving direct violation of \CPT. In addition, using this formalism we have studied three special cases too: a) SM scenario, b) SM plus direct \CP violation c) SM plus \T and \CPT violation in mixing.}

\maketitle
\flushbottom

\section{Introduction:}
\label{Intro}

\CPT invariance  is one of the most fundamental principles in Physics. It is believed that any natural process must be described by a \CPT invariant Lagrangian. According to \CPT \textit{theorem}, any quantum field theory involving point-particles in flat Minkowski space, delineated by Hermitian, local, Lorentz-invariant Lagrangian (or Hamiltonian), is certainly \CPT invariant \cite{symmetry,bigisanda}. The primary proof of this theorem was given by L\"{u}ders, Pauli and others \cite{luders1,luders2,pauli,bell,schwinger} (an updated version of this approach can be found in Ref.~\cite{winberg}) depending on the formulation of Hamiltonian (or Lagrangian) for quantum field theory. Later on the theorem was proven rigorously by Jost and others \cite{jost,bogoliubov,wightman1} in the axiomatic
formalism of quantum field theory based on the assumptions of Lorentz invariance, 
existence of unique vacuum state and weak local commutativity obeying `right' statistics. The line of proof in this approach mainly depends on Wightman axioms, Wightman functions, the Wightman reconstruction theorem and Bargmann-Hall-Wightman theorem on complex Lorentz transformations \cite{wightman1,wightman2,wightman3}. 

Nonetheless, there are various models in literature that violates \CPT by evading some of the necessary conditions of \CPT theorem. Both the approaches, mentioned above, address \CPT theorem for fundamental particles only and they fail to handle the same with QCD bound states (although Ref. \cite{sudarshan} that uses dynamical principle and variation of action methods of Schwinger \cite{schwinger} claims a third way to prove \CPT theorem incorporating bound states too). Open Bosonic strings, which are not point-like object, can go through spontaneous \CPT violation \cite{kostelecky1}. Again, violation of Lorentz symmetry may also lead to \CPT violation \cite{kostelecky2,kostelecky3,kostelecky4,kostelecky5,kostelecky6,kostelecky7,greenberg}.
 Non-trivial space-time topology could be a reason for \CPT violation, e.g. the vacuum state for a model with one of the three spatial dimensions compactified into a circle of cosmological size fails to be Lorentz invariant which in turn produce violation of \CPT symmetry \cite{klinkhamer1,klinkhamer2}. However, it should be kept in mind that the above condition is not a necessary one, for example QFT on non-commutative space-time can give rise to Lorentz invariance violating effects while conserving \CPT \cite{chaichian4,chaichian5,alvarez,jabbari}. A certain class of models can violate \CPT through non-locality too while preserving Lorentz symmetry \cite{chaichian1,chaichian2,chaichian3}. Non-point interactions, which are also a possible source of \CPT violation, emerge in some models where particle and antiparticle are both contained in the same isospin multiplet \cite{carruthers1,carruthers2,carruthers3,carruthers4,carruthers5}. \CPT violation may arise from modifications of conventional quantum mechanics due to gravitational effects \cite{hawking1,hawking2}, specially near event horizon where inaccessibility of full information leads to non-unitarity of states. Quantum-mechanical decoherence in quantum gravity could also be responsible for \CPT breaking \cite{mavromatos1,mavromatos2}. It has been shown in Refs. \cite{klinkhamer3,abers,oksak} that Abelian Chern-Simons like terms in Lagrangian as well as fields with infinite components also violate \CPT.


Given its great importance to theoretical Physics, much attention has been devoted to scrutinize the plausibility of \CPT symmetry experimentally. The observed equality between masses and life times of particle and antiparticle with striking
precision \cite{pdg}, which is a consequence of \CPT invariance, obligates us to believe that \CPT is a good symmetry of nature. But, these quantities are mainly dominated by strong or electromagnetic interactions and therefore, the possibility for existence of tiny \CPT violating effects mediated by weak interactions, which might remain undetected in direct measurements, cannot be ruled out. Apart from neutrino sector,
 mixing of neutral pseudoscalar meson ($K^0,\, D^0,\, B_d^0,\, B_s^0$) with its own antiparticle, in this regard, is a promising place to search for \CPT violating effects \cite{lavora,sanda,ellis,edw,roberts} as it is predominantly a second order electroweak phenomenon.  However, in addition to \CPT violating effects, since the most general mixing matrix involves \T and \CP violation as well, all those effects must be considered together.

In literature, there exist extensive studies on probing \T, \CP and \CPT violation using leptonic, semi-leptonic, two pseudoscalars and one pseudoscalar plus one vector decay modes of neutral pseudoscalar meson \cite{CPTVBmix_th2,Alvarez:2003kh,CPTVBmix_BaBar1,CPTVBmix_BaBar2,Alvarez:2004tj,CPTVBmix_th3,CPTVBmix_th4,expts1,CPTVBmix_th5,kundu1,kundu2,tilburg,Bernabeu:2016sgz,expts2,karan1,Botela,hflav,nogu}. But the modes with neutral pseudoscalar mesons decaying to two vectors $(P^0 \text{ or } \bar P^0\to V_1 V_2)$ are not very well assessed in light of \CPT violation. Refs. \cite{VV1,VV2,London:2003rk} consider the SM scenario (i.e. \CP violation in mixing only ) and its extension to models with \CPT conserving generic new physics effects only while probing two vectors decay modes of $B$-mesons. Howbeit, Ref. \cite{kundu3} has taken \CPT violation into account for describing the mode $B^0_s\to J/\psi\,\phi$ and Ref. \cite{kundu4} has discussed about triple products and angular observables for $B\to V_1 V_2$ decays in context of \CPT violation. Furthermore, two vectors decay modes of neutral mesons have been studied in Ref. \cite{karan2} contemplating \T, \CP and \CPT violation in mixing only. In this paper, we have extended the idea of Ref. \cite{karan2} to search for \T and \CPT violation in mixing through $P^0\to V_1 V_2$ decays using helicity-based analysis in presence of \CP violation in decay as well as in mixing. Notwithstanding, the presence of \CP violation in decay changes the scenario drastically  and complicate all the equations compared to Ref. \cite{karan2} and consequently recasting the whole approach for this analysis is essential. It should also be noted that we do not consider any specific model that might lead to \CPT violation which implies that it's a model-independent approach.

While dealing with oscillations of neutral pseudoscalar mesons $(P^0, \bar P^0)$, usually a common final state $f$, to which both $P^0$ and $\bar P^0$ can decay, is considered. When $f$ contains two vectors, there emerge three transversity amplitudes for the each of the transitions $P^0 \to f $ and $\bar P^0 \to f$ depending on the orbital angular momentum of the final state. This fact helps us to construct a large number of observables from time-dependent differential decay rates of the two modes. But, all of the observables will not be independent to each other since independent theoretical parameters are lesser in number than the observables. Therefore, various relations among observables appear automatically. These relations have already been addressed in context SM scenario in Refs. \cite{VV2,London:2003rk,karan2}, where first two references consider two vector decay modes of $B^0_d$ only with vanishing width deference between the physical states. Moreover, Ref. \cite{karan2} talks about these relations in the presence of \T, \CP and \CPT violation in mixing. In this paper we advance one step further by exploring these relations in the presence of \CP violation in decay in addition to \T, \CP and \CPT violating effects in mixing. These new relations will break down only if any \CPT violating effect is present in decay itself. Furthermore, we have used our formalism to study the relations among observables for SM scenario, SM plus direct \CP violation case and SM plus indirect violation of \T and \CPT scenario, which are three special cases of our picture.

The paper is organized as follows. In the next section (Sec. \ref{sec:Theory}), we briefly describe the theoretical formalism for \CPT violation in $P^0-\bar P^0$ mixing and express the time dependent differential decay rates of $P^0$ and $\bar P^0$ in terms of the mixing parameters. In Sec. \ref{sec:Observables}, we construct helicity-dependent observables from the differential decay rates and express the ``dummy observables" in terms of them as well as small \T and \CPT violating parameters. Sec. \ref{sec:Solution} deals with solving unknown theoretical parameters in terms of observables. The relations among observables in present scenario have been established in Sec. \ref{sec:ObservableRelations}. In Sec. \ref{sec:Splcs}, we use this formalism to find the relations among observables for three special cases: a) SM scenario (\CP violation in mixing only), b) SM plus direct \CP violation, c) SM plus \T and \CPT violation in mixing. The phenomenological aspects have been discussed in Sec. \ref{sec:Phenomenology} and finally, we summarize and conclude in Sec. \ref{sec:Conclusion}.

\section{Theoretical Framework:}
\label{sec:Theory}

Let us first briefly review the most general formalism incorporating  \CPT and  \T violation for $P^0-\bar P^0 $mixing, which has already been discussed in Ref. \cite{,bigisanda,karan1,karan2}. In the flavour basis $(P^0,\bar P^0)$, the mixing Hamiltonian can be expressed in terms of two $2\times2$ Hermitian matrices, namely mass-matrix $\mathbf M$ and decay-matrix $\mathbf \Gamma$, as $\mathbf {M}-(i/2) \mathbf \Gamma$. Since three Pauli matrices $\bm\sigma_j$ along with identity matrix $\mathbf I$ constitute a complete set of bases spanning the whole vector-space of $2\times2$ matrices, one can write:
\begin{equation}
\label{hamiltonian}
\mathbf M-\frac{i}{2}\mathbf\Gamma=E\sin\theta\cos\phi\,\bm\sigma_1+E\sin\theta\sin\phi\,\bm\sigma_2+E\cos\theta\,\bm\sigma_3-iD\,\mathbf I
\end{equation}
where, $E,\theta,\phi$ and $D$ are complex entities in general. Comparing both sides of this equation, we obtain:
\begin{equation}
\label{EDdefs}
\begin{split}
&D = \frac{i}{2} (M_{11} + M_{22}) + \frac{1}{4} (\Gamma_{11} + \Gamma_{22}) ~,\\
&E\cos\theta = \frac{1}{2} \, (M_{11} - M_{22}) - \frac{i}{4} (\Gamma_{11} - \Gamma_{22}) ~,\\
&E\sin\theta\cos\phi = {\rm Re} \, M_{12}-\frac{i}{2}{\rm Re} \, \Gamma_{12} ~,\\
&E\sin\theta\sin\phi = -{\rm Im} \, M_{12} + \frac{i}{2} {\rm Im} \, \Gamma_{12} ~.
\end{split}
\end{equation}
where $M_{ij}$ and $\Gamma_{ij}$ are $(i,j)$-th elements of $\mathbf M$ and $\mathbf \Gamma$ matrices respectively.

The mass eigenstates  or physical states $\ket{P_L}$ and $\ket{P_H}$ are th eigenvectors of the mixing Hamiltonian $\mathbf {M}-(i/2) \mathbf \Gamma$  and they can be expressed as linear combinations of the flavour eigenstates ($\ket{P^0}$ and $\ket{\bar P^0}$) as follows:
\begin{equation}
\label{MState}
\begin{split}
\ket{P_L}= p_1 \ket{P^0}+q_1 \ket{\bar{P}^0}, \qquad \quad \ket{P_H}= p_2 \ket{P^0}-q_2 \ket{\bar{P}^0},
\end{split}
\end{equation}
where $p_1=N_1\cos{\frac{\theta}{2}},\ q_1=N_1\,e^{i\phi}\sin{\frac{\theta}{2}},\ p_2=N_2\sin{\frac{\theta}{2}}$,  $q_2=N_2\,e^{i\phi}\cos{\frac{\theta}{2}}$ with $N_1, N_2$ being two normalization factors and the $L$,$H$ tags indicating light and heavy physical states, respectively. Since, the physical states, as given by Eq.~\eqref{MState}, depend only on the  complex parameters $\theta$ and $\phi$, they are called the mixing parameters for \PPbr system. It should be noted that the physical states are not orthogonal in general since the mixing matrix is non-Hermitian. 

The time evolution of flavour states  ($\ket{P^0} \equiv \ket{P^0(t=0)}$ and $\ket{{\bar P}^0} \equiv \ket{{\bar	P}^0(t=0)}$) is given by:
\begin{equation}
\label{TState}
\begin{split}
&\ket{P^0(t)}=  h_{+} \ket{P^0} + h_{-}\cos{\theta}\ket{P^0}  + h_{-} e^{i \phi} \sin{\theta} \ket{\bar P^0}, \\[2mm]
&\ket{\bar P^0(t)}= h_{+} \ket{\bar P^0} - h_{-}\cos{\theta}\ket{\bar P^0}  + h_{-} e^{-i \phi} \sin{\theta} \ket{P^0},\\[2mm]
\text{where, \hspace*{3mm}}\qquad  &h_{\pm}= e^{-i\big(M-i \frac{\Gamma}{2}\big)t}\bigg[\frac{e^{i\big(\Delta{M}-i\frac{\Delta\Gamma }{2}\big)\frac{t}{2}}\pm e^{-i\big(\Delta{M}-i\frac{\Delta\Gamma }{2}\big)\frac{t}{2}}}{2}\bigg].\qquad\qquad
\end{split}
\end{equation}
Here, $M=(M_H + M_L)/2$, $\Delta M=M_H-M_L$, $\Gamma=(\Gamma_H + \Gamma_L)/2$ and $\Delta\Gamma=\Gamma_H-\Gamma_L$ with $M_{L,H}$ and $\Gamma_{L,H}$ to be masses and decay widths of the light and heavy mass eigenstates respectively which can be found from the eigenvalues of the mixing Hamiltonian and measured directly in experiments. 

Let us now consider a final state $f$ to which both $P^0$ and $\bar P^0$ can decay. Using Eq.~\eqref{TState}, the time dependent decay amplitudes for the neutral mesons are given by:
\begin{equation}
\label{TAmplitude}
\begin{split}
\mathcal{A}mp(P^0(t)\rightarrow f)=& h_{+} \mathcal{A}_{f} + h_{-} \cos{\theta} \mathcal{A}_{f} + h_{-} e^{i \phi} \sin{\theta} \bar{\mathcal{A}}_{f},\\[2mm]
\mathcal{A}mp(\bar P^0(t)\rightarrow f)=& h_{+} \bar{\mathcal{A}}_{f} - h_{-} \cos{\theta} \bar{\mathcal{A}}_{f} + h_{-} e^{-i \phi} \sin{\theta} \mathcal{A}_{f}, 
\end{split}
\end{equation} 
where $\mathcal{A}_f= \bra{f}\mathcal{H}_{\Delta F=1}\ket{P^0} \text{ and }\bar{\mathcal{A}}_{f}=\bra{f}\mathcal{H}_{\Delta F=1}\ket{\bar P^0}$ with $\mathcal{H}_{\Delta F=1}$ indicating the Hamiltonian related to the transition from flavour states to $f$. Therefore, incorporating the mixing, the time dependent decay rates $\Gamma(P^0(t)\rightarrow f)$ and $\Gamma(\bar P^0(t)\rightarrow f)$ can be expressed as:
\small
\begin{eqnarray}
\label{TGamma}
\frac{d\Gamma}{dt}(P^0(t)\to   f) &=& \frac12 e^{-\Gamma t} 
\left[
\sinh\left(\Delta\Gamma t/2\right)  \left\{ 2\text{Re} 
\left( \cos\theta |\mathcal{A}_{f}|^2 +e^{i \phi} \sin\theta 
\mathcal{A}_{f}^{*} \bar{\mathcal{A}}_{f} \right) \right\}
\right. \nn \\
&& \hskip-1truein  +~\cosh\left(\Delta\Gamma t/2\right) \left\{ |\mathcal{A}_f |^2 + |\cos\theta |^2|\mathcal{A}_f |^2 + |e^{i \phi}\sin\theta |^2|\bar{\mathcal{A}}_{f} |^2 
+ 2 \text{Re}\left(e^{i \phi} \cos\theta^{*} \sin\theta 
\mathcal{A}_{f}^{*} \bar{\mathcal{A}}_{f} \right) \right\} \nn\\
&& \hskip-1truein  +~\cos(\Delta M t) \left\{ |\mathcal{A}_f |^2 - |\cos\theta |^2|\mathcal{A}_f |^2 - |e^{i \phi}\sin\theta |^2|\bar{\mathcal{A}}_{f} |^2 - 2 \text{Re}\left(e^{i \phi} \cos\theta^{*} \sin\theta 
\mathcal{A}_{f}^{*} \bar{\mathcal{A}}_{f} \right)
\right\} \nn\\
&& \left. -~\sin(\Delta Mt) \left\{ 2 \text{Im} \left( \cos\theta |\mathcal{A}_{f}|^2 + e^{i \phi} \sin\theta \mathcal{A}_{f}^{*}\bar{\mathcal{A}}_{f}
\right) \right\}
\right] ~,
\end{eqnarray}
\begin{eqnarray}
\label{TGammaPar}
\frac{d\Gamma}{dt}({\bar P}^0(t)\to  f) &=& \frac12 e^{-\Gamma t} \left[
\sinh\left(\Delta\Gamma t/2\right) \left\{ 2\text{Re} 
\left( -\cos\theta^{*} |\bar{\mathcal{A}}_{f}|^2 + e^{i \phi^*} 
\sin\theta^{*} \mathcal{A}_{f}^{*} \bar{\mathcal{A}}_{f} \right) \right\} \right. \nn \\
&& \hskip-1truein +~\cosh\left(\Delta\Gamma t/2\right) \left\{ |\bar{\mathcal{A}}_{f} |^2 + |\cos\theta |^2|\bar{\mathcal{A}}_{f} |^2 + | e^{-i \phi}\sin\theta |^2|\mathcal{A}_{f} |^2 
- 2 \text{Re}\left( e^{i \phi^*} \cos\theta \sin\theta^{*} \mathcal{A}_{f}^{*} \bar{\mathcal{A}}_{f} \right)
\right\} \nn\\
&& \hskip-1truein +~\cos(\Delta M t) \left\{ |\bar{\mathcal{A}}_{f} |^2 - |\cos\theta |^2|\bar{\mathcal{A}}_{f} |^2 - | e^{-i \phi}\sin\theta |^2|\mathcal{A}_{f} |^2 + 2 \text{Re}\left( e^{i \phi^*} \cos\theta \sin\theta^{*} 
\mathcal{A}_{f}^{*} \bar{\mathcal{A}}_{f} \right)
\right\} \nn\\
&& \left. +~\sin(\Delta Mt) \left\{ 2 \text{Im} \left( -\cos\theta^{*}  
|\bar{\mathcal{A}}_{f}|^2 +  e^{i \phi^*} \sin\theta^{*} 
\mathcal{A}_{f}^{*} \bar{\mathcal{A}}_{f}
\right) \right\} \right] ~.
\end{eqnarray}
\normalsize

\section{Observables:}
\label{sec:Observables}

\subsection{T and CPT violating parameters:}
\label{T-CPTparameters}

The properties of $\mathbf M$ and $\mathbf \Gamma$ matrices in light of \T and \CPT symmetry has been discussed in Ref. \cite{lee}. First, if \CPT invariance holds, then, independently of \T symmetry \cite{karan1,karan2},
\begin{equation}
\label{CPTgood}
M_{11} = M_{22} ~,~~ \Gamma_{11} = \Gamma_{22} \quad \Longrightarrow \theta = \frac{\pi}{2} \quad \text{(Using Eq. \eqref{EDdefs})}.
\end{equation}
Secondly, if \T invariance holds, then, independently of \CPT symmetry \cite{karan1,karan2},
\begin{equation}
\label{Tgood}
\frac{\Gamma_{12}^*}{\Gamma_{12}} = \frac{M_{12}^*}{M_{12}}
\Longrightarrow {\rm Im} \, \phi = 0 ~\quad \text{(Using Eq. \eqref{EDdefs})}.
\end{equation}
Hence, incorporating \T, \CP and \CPT violation in  \PPbr mixing, we parametrize $\theta$ and $\phi$ as \cite{karan1,karan2}:
\begin{equation}
\label{epsdef}
\theta=\frac{\pi}{2}+\epsilon_1+i\epsilon_2\; \text{ and }\; \phi=-2\beta+i\epsilon_3
\end{equation}
where, $\beta$ is the \CP violating weak mixing phase, $\epsilon_1\text{ and }\epsilon_2$ are \CPT violating parameters and $\epsilon_3$ is \T violating parameter other than \CP violation. The notation of Belle, BaBar and LHCb  collaborations \cite{CPTVBmix_BaBar1,CPTVBmix_BaBar2,expts1,expts2} is a bit different from ours; however, the two notations are related to each other by the following transformation \cite{karan1,karan2}:
\begin{equation}
\label{epsdefs}
\begin{split}
&\hspace*{-1.2cm}\cos\theta \leftrightarrow  -z ~,~~ \sin\theta \leftrightarrow \sqrt{1-z^2} ~,~~ e^{i\phi} \leftrightarrow \frac{q}{p} ~,\\
\text{or, equivalently: }&\quad \epsilon_1 = {\rm Re}(z) ~,~~ \epsilon_2 = {\rm Im}(z) ~,~~ \epsilon_3 = 1-\Big|\frac{q}{p}\Big| ~.
\end{split}
\end{equation}

\subsection{Decay rates and observables:}
\label{DecayRates}

Any state consisting of two vectors can have three different values for orbital angular momentum quantum number $\{0,1,2\}$ which correspond to the polarization states $\{0,\perp,\parallel\}$, respectively. Since \CPT violation in decay has not been considered, the decay amplitudes for modes and conjugate modes can be expressed in terms of transversity amplitudes as \cite{VV1,VV2,London:2003rk,kundu4,karan2}:
\begin{equation}
\label{VVAmplitude}
\begin{split}
\mathcal{A}_{f}(P^0\rightarrow V_{1}V_{2}) =& A_{0} g_{0} + A_{\pl} g_{\pl} + iA_{\perp} g_{\perp}, \\
\widebar{\mathcal{A}_{f}}(\bar P^0\rightarrow V_{1}V_{2}) =& \bar A_{0} g_{0} + \bar A_{\pl} g_{\pl} - i\bar{A}_{\perp} g_{\perp}~. 
\end{split}                          
\end{equation}
where the factors $g_\lambda$ with $\lambda\in\{0,\parallel,\perp\}$ are the coefficients of transversity amplitudes ($A_\lambda$ or $\bar A_\lambda$) in linear polarization basis and depend only on the kinematic angles~\cite{Sinha:1997zu}.

Now, using Eq.~\eqref{TGamma}-\eqref{VVAmplitude}, the time-dependent decay rates for $P^0(t)\rightarrow V_1V_2$ and $\bar{P}^0(t)\rightarrow V_1V_2$ modes can be written as~\cite{VV1,VV2,London:2003rk, kundu3, kundu4,karan2}:
\small
\begin{align}
\label{VVDecayRate}
\begin{split}
&\frac{d\Gamma}{dt}(P^0(t)\to V_1V_2\big) = \\
&e^{-\Gamma t}\sum_{\lambda\leq\sigma}\Bigg[\Lambda_{\lambda\sigma}\cosh\Big(\frac{\Delta\Gamma t}{2}\Big)+\eta_{\lambda\sigma}^{}\sinh\Big(\frac{\Delta\Gamma t}{2}\Big)+\Sigma_{\lambda\sigma}\cos\big(\Delta M t\big)-\rho_{\lambda\sigma}^{}\sin\big(\Delta M t\big)\Bigg]g_\lambda g_\sigma~,
\end{split} \\
\label{ConjVVDecay}
\begin{split}
&\frac{d\Gamma}{dt}(\bar{P}^0(t)\to V_1V_2\big) = \\
&e^{-\Gamma t}\sum_{\lambda\leq\sigma}\Bigg[\bar{\Lambda}_{\lambda\sigma}\cosh\Big(\frac{\Delta\Gamma t}{2}\Big)+\bar{\eta}_{\lambda\sigma}^{}\sinh\Big(\frac{\Delta\Gamma t}{2}\Big)+\bar{\Sigma}_{\lambda\sigma}\cos\big(\Delta M t\big)+\bar{\rho}_{\lambda\sigma}^{}\sin\big(\Delta M t\big)\Bigg]g_\lambda g_\sigma~.
\end{split}
\end{align}
\normalsize
where both $\lambda$ and $\sigma$ take the value $\{0,\parallel,\perp\}$. It is important to note that in the entire paper we have considered a particular ordering for the combination $\lambda\sigma$ with $\lambda\neq\sigma$ and they are \{$\perp 0$, $\perp \parallel$, $\parallel  0$\}. For defining the observables, we have taken the convention of Ref. \cite{karan2}. On the other hand, Refs. \cite{London:2003rk,VV1,VV2} use a bit different notations involving some additional negative signs. Hence, several inferences of our paper may differ from their results by some signs only; however, all the outcomes of our paper are self-consistent.

 Now, we see from Eq.~\eqref{VVDecayRate} that for each of the helicity combinations, there are four types of observables $(\Lambda_{\ls},\eta_{\ls},\Sigma_{\ls}, \rho_{\ls})$ and six such helicity combinations are possible. Hence, we get total 24 observables for $P^0(t)\rightarrow V_1V_2$ mode. Similarly, there will be 24 different observables $(\bar{\Lambda}_{\ls},\bar{\eta}_{\ls},\bar{\Sigma}_{\ls}, \bar{\rho}_{\ls})$ for $\bar{P}^0((t)\rightarrow V_1V_2$ mode too, as shown in Eq.~\eqref{ConjVVDecay}. These observables can be measured by performing a time dependent angular analysis of $P^0(t)\rightarrow V_1V_2$ and $\bar P^0(t)\rightarrow V_1V_2$ \cite{VV1,VV2,London:2003rk}. The procedure described in Ref. \cite{kundu4} can be helpful in this regard.  On the other hand, probing polarizations of the final state particles may also aid in measurement of these observables. It should be noticed that Ref.~\cite{VV1,VV2,London:2003rk} did not consider  $\sinh\Big(\frac{\Delta\Gamma t}{2}\Big)$ terms in the decays of $B^0_d$ and $\bar B^0_d$ since $\Delta\Gamma$ is consistent with zero \cite{pdg}.
In that case, $\eta_{\lambda\sigma}$ and $\bar\eta_{\lambda\sigma}$ remain undetermined and one should work with remaining $(18+18)=36$ observables only for a mode and its conjugate mode. However, we have kept all the terms in our analysis since a general scenario has been considered here.  

\subsection{Parametric expansion of observables:}
Comparing Eqs.~\eqref{TGamma} and \eqref{TGammaPar} to Eqs.~\eqref{VVDecayRate} and \eqref{ConjVVDecay} one can easily infer that all of the observables will be functions of the complex quantities $\theta$ and $\phi$. As \T and \CPT violations are expected to be very small \cite{CPTVBmix_BaBar1,CPTVBmix_BaBar2,expts1,expts2,hflav,ktev,focus}, we can expand all the observables in terms of $\epsilon_j$ ($j\in\{1,2,3\}$)  keeping up to the linear orders. However, following Refs. \cite{VV2,London:2003rk}, if we divide all the transversity amplitudes into \CP conserving and \CP violating parts, it would become very complicated to handle all the unknown parameters. So, we implement a new method to reduce the complexity. After substitution of Eqs. \eqref{epsdef} and \eqref{VVAmplitude} into Eqs. \eqref{TGamma} and \eqref{TGammaPar}, while expanding the differential decay rates in terms of $\epsilon_j$  ($j\in\{1,2,3\}$), we find that only twenty four combinations of helicity amplitudes $A_\lambda$ and $\bar{A}_\lambda$ appear as the coefficients of $\epsilon_j$. Denoting $\xi=e^{-2i\beta}$, we define these twenty four combinations as follows:

\begin{minipage}{0.49\textwidth}

	\begin{equation}
	\begin{split}
	\hspace*{0.5cm}	&\Lambda'_{\lambda\lambda} = \frac{1}{2} (|A_\lambda|^2 + |\bar A_\lambda|^2 )\,, \\  &\Lambda'_{\perp i} = -\text {Im }(A_{\perp}A_i^*-\bar A_{\perp}\bar A_i^*)\,, \\  &
	\Lambda'_{\parallel 0} = \text {Re }(A_{\parallel}A_0^* + \bar A_{\parallel}\bar A_0^*)\,,  \\  &
	\rho'_{ii} = \text {Im }(\xi A_{i}^* \bar A_{i})\,,  \\  &
	\rho'_{\perp\perp} =- \text {Im }(\xi A_{\perp}^* \bar A_{\perp})\,,  \\  &
	\rho'_{\perp i} = -\text {Re }\big[\xi(A_{\perp}^*\bar A_i +A_{i}^*\bar A_{\perp} )\big]\,,\\  &
	\rho'_{\parallel 0} = \text {Im }\big[\xi(A_{\parallel}^*\bar A_0 +A_{0}^*\bar A_{\parallel} )\big]\,,\nonumber
	\end{split}
	\end{equation}
\end{minipage}
\hfil
\hspace*{-1cm}\begin{minipage}{0.51\textwidth}
	\begin{equation}
	\label{dummy obs}
	\begin{split}
	& \Sigma'_{\lambda\lambda} = \frac{1}{2} (|A_\lambda|^2 - |\bar A_\lambda|^2 )\,,  \\  &	\Sigma'_{\perp i} = -\text {Im }(A_{\perp}A_i^* + \bar A_{\perp}\bar A_i^*)\,, \\  &
 \Sigma'_{\parallel 0} = \text {Re }(A_{\parallel}A_0^* - \bar A_{\parallel}\bar A_0^*)\,,  \\  &
\eta'_{ii} = \text {Re }(\xi A_{i}^* \bar A_{i})\,,\\&
 \eta'_{\perp\perp} = -\text {Re }(\xi A_{\perp}^* \bar A_{\perp})\,,  \\  &
\eta'_{\perp i} = \text {Im }\big[\xi(A_{\perp}^* \bar A_i + A_{i}^* \bar A_{\perp})\big ]\,,\\  &
\eta'_{\parallel 0} = \text {Re }\big[\xi(A_{\parallel}^* \bar A_0 + A_{0}^* \bar A_{\parallel} )\big]\,,
	\end{split}
	\end{equation}
\end{minipage}

\vspace*{3mm}
 \noindent where, $i\in\{0,\parallel\}$ and $\lambda\in\{0,\parallel,\perp\}$. It should be noted that the quantities, mentioned above, which we name as ``dummy-observables", are not observables, in general; rather they are some theoretical tools for our convenience. Now, using Eqs. \eqref{VVDecayRate} and \eqref{ConjVVDecay} one can express the actual observables in terms of the dummy-observables as well as the \T and \CPT violating parameters $\epsilon_j$. For our purpose, we invert those equations and express dummy-observables as functions of the original ones keeping only the linear orders in $\epsilon_{j}$ as follows:

\small
\begin{minipage}[t]{0.5\textwidth}
	\begin{equation}
	\begin{split}
	&\hspace*{-1cm}\Lambda_{ii}^{\prime} = \epsilon_1 \eta_{ii} + (1+\epsilon_3) \Lambda_{ii} + \epsilon_2 \rho_{ii} - \epsilon_3 \Sigma_{ii} \,,\\
	&\hspace*{-1cm}\Lambda_{\pp}^{\prime} = \epsilon_1 \eta_{\pp} + (1+\epsilon_3) \Lambda_{\pp} + \epsilon_2 \rho_{\pp} - \epsilon_3 \Sigma_{\pp} \,,\\
	&\hspace*{-1cm}\Lambda_{\perp i}^{\prime} = \epsilon_1 \eta_{\perp i} + (1+\epsilon_3) \Lambda_{\perp i} + \epsilon_2 \rho_{\perp i} - \epsilon_3 \Sigma_{\perp i} \,,\\
	&\hspace*{-1cm}\Lambda_{\parallel 0}^{\prime} = \epsilon_1 \eta_{\parallel 0} + (1+\epsilon_3) \Lambda_{\parallel 0} + \epsilon_2 \rho_{\parallel 0} - \epsilon_3 \Sigma_{\parallel 0} \,,\\
	&\hspace*{-1cm}\Sigma_{ii}^{\prime} = -\epsilon_1 \eta_{ii} - \epsilon_3 \Lambda_{ii} - \epsilon_2 \rho_{ii} + (1+\epsilon_3) \Sigma_{ii} \,,\\
	&\hspace*{-1cm}\Sigma_{\pp}^{\prime} = -\epsilon_1 \eta_{\pp} - \epsilon_3 \Lambda_{\pp} - \epsilon_2 \rho_{\pp} + (1+\epsilon_3) \Sigma_{\pp} \,,\\
	&\hspace*{-1cm}\Sigma_{\perp i}^{\prime} = -\epsilon_1 \eta_{\perp i} - \epsilon_3 \Lambda_{\perp i} - \epsilon_2 \rho_{\perp i} + (1+\epsilon_3) \Sigma_{\perp i} \,,\\
	&\hspace*{-1cm}\Sigma_{\parallel 0}^{\prime} = -\epsilon_1 \eta_{\parallel 0} - \epsilon_3 \Lambda_{\parallel 0} - \epsilon_2 \rho_{\parallel 0} + (1+\epsilon_3) \Sigma_{\parallel 0} \,,  \nonumber\\
	\end{split}
	\end{equation}	
\end{minipage}
\hfill
\begin{minipage}[t]{0.42\textwidth}
	\begin{equation}
		\label{prime_to_mode}
		\begin{split}
		&\hspace*{-0.5cm}\eta_{ii}^{\prime} = (1+\epsilon_3)\eta_{ii} + \epsilon_1 \Lambda_{ii} + \epsilon_1 \Sigma_{ii} \,,\\
		&\hspace*{-0.5cm}\eta_{\pp}^{\prime} = (1+\epsilon_3)\eta_{\pp} + \epsilon_1 \Lambda_{\pp} + \epsilon_1 \Sigma_{\pp} \,,\\
		&\hspace*{-0.5cm}\eta_{\perp i}^{\prime} = (1+\epsilon_3)\eta_{\perp i} + \epsilon_1 \Lambda_{\perp i} + \epsilon_1 \Sigma_{\perp i} \,,\\
		&\hspace*{-0.5cm}\eta_{\parallel 0}^{\prime} = (1+\epsilon_3)\eta_{\parallel 0} + \epsilon_1 \Lambda_{\parallel 0} + \epsilon_1 \Sigma_{\parallel 0} \,,\\
		&\hspace*{-0.5cm}\rho_{ii}^{\prime} = \epsilon_2 \Lambda_{ii} + (1+\epsilon_3) \rho_{ii} + \epsilon_2 \Sigma_{ii} \,,\\
		&\hspace*{-0.5cm}\rho_{\pp}^{\prime} = \epsilon_2 \Lambda_{\pp} + (1+\epsilon_3) \rho_{\pp} + \epsilon_2 \Sigma_{\pp} \,,\\
		&\hspace*{-0.5cm}\rho_{\perp i}^{\prime} = \epsilon_2 \Lambda_{\perp i} + (1+\epsilon_3) \rho_{\perp i} + \epsilon_2 \Sigma_{\perp i} \,,\\
		&\hspace*{-0.5cm}\rho_{\parallel 0}^{\prime} = \epsilon_2 \Lambda_{\parallel 0} + (1+\epsilon_3) \rho_{\parallel 0} + \epsilon_2 \Sigma_{\parallel 0} \,,
	\end{split}
\end{equation}
\end{minipage}

\normalsize
\noindent where, $i\in\{0,\parallel\}$. It is evident from above relations that the dummy-observables become original observable only when there is no \T and \CPT violation in mixing. Now, we use the same trick for the observables of conjugate mode too and taking $i\in\{0,\parallel\}$ rewrite the dummy-observables in terms of them as follows:

\small
\begin{minipage}[t]{0.5\textwidth}
	\begin{equation}
	\begin{split}
	&\hspace*{-1cm}\Lambda_{ii}^{\prime} = -\epsilon_1 \bar \eta_{ii} + (1-\epsilon_3) \bar\Lambda_{ii} + \epsilon_2 \bar\rho_{ii} + \epsilon_3 \bar\Sigma_{ii} \,,\\ & \hspace*{-1cm}
	\Lambda_{\perp\perp}^{\prime} = -\epsilon_1 \bar\eta_{\perp\perp} + (1-\epsilon_3) \bar\Lambda_{\perp\perp} +\epsilon_2 \bar\rho_{\perp\perp} + \epsilon_3 \bar\Sigma_{\perp\perp} \,,\\ & \hspace*{-1cm}
	\Lambda_{\perp i}^{\prime} = -\epsilon_1 \bar\eta_{\perp i} + (1-\epsilon_3)\bar \Lambda_{\perp i} + \epsilon_2 \bar\rho_{\perp i} + \epsilon_3 \bar\Sigma_{\perp i} \,,\\ & \hspace*{-1cm}
	\Lambda_{\parallel 0}^{\prime} = -\epsilon_1 \bar\eta_{\parallel 0} + (1-\epsilon_3) \bar\Lambda_{\parallel 0} + \epsilon_2 \bar\rho_{\parallel 0} + \epsilon_3 \bar\Sigma_{\parallel 0} \,,\\ & \hspace*{-1cm}
	\Sigma_{ii}^{\prime} = -\epsilon_1\bar \eta_{ii} - \epsilon_3\bar \Lambda_{ii} + \epsilon_2 \bar\rho_{ii} - (1-\epsilon_3)\bar \Sigma_{ii} \,,\\ & \hspace*{-1cm}
	\Sigma_{\perp\perp}^{\prime} = -\epsilon_1 \bar\eta_{\perp\perp} - \epsilon_3 \bar\Lambda_{\perp\perp} + \epsilon_2 \bar\rho_{\perp\perp} - (1-\epsilon_3) \bar\Sigma_{\perp\perp} \,,\\ & \hspace*{-1cm}
	\Sigma_{\perp i}^{\prime} = -\epsilon_1 \bar\eta_{\perp i} - \epsilon_3 \bar\Lambda_{\perp i} + \epsilon_2 \bar\rho_{\perp i} - (1-\epsilon_3) \bar\Sigma_{\perp i} \,,\\ & \hspace*{-1cm}
	\Sigma_{\parallel 0}^{\prime} = -\epsilon_1\bar \eta_{\parallel 0} - \epsilon_3 \bar\Lambda_{\parallel 0} + \epsilon_2\bar \rho_{\parallel 0} - (1-\epsilon_3) \bar\Sigma_{\parallel 0} \,,\nonumber
		\end{split}
		\end{equation}	
	\end{minipage}
	\hfill
	\begin{minipage}[t]{0.42\textwidth}
		\begin{equation}
		\label{prime_to_conj}
		\begin{split}
		&\hspace*{-0.5cm}\eta_{ii}^{\prime} = (1-\epsilon_3)\bar\eta_{ii} - \epsilon_1\bar \Lambda_{ii} - \epsilon_1 \bar\Sigma_{ii} \,,\\ & \hspace*{-0.5cm}
		\eta_{\perp\perp}^{\prime} = (1-\epsilon_3)\bar\eta_{\perp\perp} - \epsilon_1 \bar\Lambda_{\perp\perp} - \epsilon_1\bar \Sigma_{\perp\perp} \,,\\ & \hspace*{-0.5cm}
		\eta_{\perp i}^{\prime} = (1-\epsilon_3)\bar\eta_{\perp i} - \epsilon_1 \bar\Lambda_{\perp i} - \epsilon_1\bar \Sigma_{\perp i} \,,\\ & \hspace*{-0.5cm}
		\eta_{\parallel 0}^{\prime} = (1-\epsilon_3)\bar\eta_{\parallel 0} - \epsilon_1 \bar\Lambda_{\parallel 0} - \epsilon_1\bar \Sigma_{\parallel 0} \,,\\ & \hspace*{-0.5cm}
		\rho_{ii}^{\prime} = \epsilon_2 \bar\Lambda_{ii} + (1-\epsilon_3) \bar\rho_{ii} + \epsilon_2 \bar\Sigma_{ii} \,,\\ & \hspace*{-0.5cm}
		\rho_{\perp\perp}^{\prime} = \epsilon_2\bar \Lambda_{\perp\perp} + (1-\epsilon_3) \bar\rho_{\perp\perp} + \epsilon_2\bar \Sigma_{\perp\perp} \,,\\ & \hspace*{-0.5cm}
		\rho_{\perp i}^{\prime} = \epsilon_2\bar \Lambda_{\perp i} + (1-\epsilon_3) \bar\rho_{\perp i} + \epsilon_2 \bar\Sigma_{\perp i} \,,\\ & \hspace*{-0.5cm}
		\rho_{\parallel 0}^{\prime} = \epsilon_2\bar \Lambda_{\parallel 0} + (1-\epsilon_3) \bar\rho_{\parallel 0} + \epsilon_2\bar \Sigma_{\parallel 0} \,.
			\end{split}
			\end{equation}
		\end{minipage}

\normalsize

\section{Solutions of the theoretical parameters:}
\label{sec:Solution}

In this section, we discuss how to solve for the unknown theoretical quantities in terms of observables. These theoretical  parameters are six helicity amplitudes ($A_\lambda$, $\bar{A}_\lambda$ with $\lambda\in\{0,\parallel,\perp\}$), which are complex entities and  three parameters $\epsilon_j$  ($j\in\{1,2,3\}$) related to \T and \CPT violation in mixing. It should be noted that the \CP violating weak phase $\beta$ cannot be probed directly in the presence of direct \CP violation; it can only be measured if there is no \CP violation in decay itself. Now, it is impossible to measure the absolute phases for all the helicity amplitudes; rather relative phases can be estimated. Hence, we define the following quantities that indicate the relative phases of five transversity amplitudes with respect to $A_\perp$:
\begin{equation}
\label{def:Omega}
\Omega_i=\text{Arg}[A_i]-\text{Arg}[A_\perp], \quad \text{and}\quad\bar\Omega_\lambda=\text{Arg}[\bar A_\lambda]-\text{Arg}[A_\perp]
\end{equation}
where, `Arg' implies argument of a complex number, $i\in\{0,\parallel\}$ and $\lambda\in\{0,\parallel,\perp\}$. Thus we have to solve for fourteen unknown parameters (three of $|A_\lambda|$, three of $|\bar A_\lambda|$, three of $\epsilon_j$, two of $\Omega_i$ and three of $\bar \Omega_\lambda$).

For convenience, we define nine angular quantities as follows:
\begin{equation}
\label{angular}
\begin{split}
\omega_{\lambda\sigma}=\text{Arg}&[A_\lambda]-\text{Arg}[A_\sigma],\qquad\quad\bar\omega_{\lambda\sigma}=\text{Arg}[\bar A_\lambda]-\text{Arg}[\bar A_\sigma],\\ &\varphi^{meas}_\lambda=-2\beta+\text{Arg}[\bar A_\lambda]-\text{Arg}[A_\lambda]\,.
\end{split}
\end{equation}
where $(\lambda,\sigma)\in\{0,\parallel,\perp\}$. As mentioned earlier, we will consider the combination $\lambda\sigma$ to be one of \{$\perp 0$, $\perp \parallel$, $\parallel  0$\} only; one should not be bothered about reverse ordering. Now, instead of the five relative phases of the transversity amplitudes we use five of the above-defined angular entities (three of $\varphi^{meas}_\lambda$ and two of $\omega_{\perp i}$) as our unknown parameters to solve for. The rest four angular quantities in Eq. \eqref{angular} will be used later in order to find relations among various observables. The relative phases of helicity amplitudes can easily be expressed in terms of the five angular entities mentioned above in the following way: 
\begin{equation}
\label{Omega}
\Omega_i=-\omega_{\perp i}\,,\qquad
\bar\Omega_i=\varphi^{meas}_i+2\beta-\omega_{\perp i}\,,\qquad \bar\Omega_\perp=\varphi^{meas}_\perp+2\beta\,,
\end{equation}
where, $i\in\{0,\parallel\}$. Thus the fourteen theoretical parameters that we are going to solve are three of $|A_\lambda|$, three of $|\bar A_\lambda|$, three of $\epsilon_j$, two of $\omega_{\perp i}$ and three of $\varphi^{meas}_\lambda$.

The modulus of helicity amplitudes are given by:
\begin{equation}
\label{modA}
|A_\lambda|=\sqrt{\Lambda_{\lambda\lambda}+\Sigma_{\lambda\lambda}}\qquad\text{and}\qquad
|\bar A_\lambda|=\sqrt{\bar\Lambda_{\lambda\lambda}+\bar\Sigma_{\lambda\lambda}}\;.
\end{equation}
The value of $\sin\Phi_\lambda^{meas}$ can be found by solving the following cubic equation:
\begin{equation}
\label{sinphi}
\begin{split}
\sin^3\Phi_\lambda^{meas}-\Big(\frac{\rho^r_{\lambda\lambda}+\bar\rho^r_{\lambda\lambda}}{\sqrt{1-C_\lambda^2}}\Big)\, \sin^2\Phi_\lambda^{meas}+\Big(\frac{C_\lambda^2-\Sigma^r_{\lambda\lambda}-\bar\Sigma^r_{\lambda\lambda}}{1-C_\lambda^2}\Big)\,\sin\Phi_\lambda^{meas}\qquad\qquad\quad\\
+\frac{C_\lambda}{\sqrt{1-C_\lambda^2}}\Big[\frac{\rho^r_{\lambda\lambda}}{1+C_\lambda}-\frac{\bar\rho^r_{\lambda\lambda}}{1-C_\lambda}\Big]=0\;,
\end{split}
\end{equation}
where, $\Phi_i^{meas}=\varphi^{meas}_i$ and $\Phi_\perp^{meas}=-\varphi^{meas}_\perp$ in the above expression. 

\noindent The quantities with superscript `r' and $C_\lambda$ are defined as:
\begin{equation}
\label{XandC}
\begin{split}
C_\lambda=\frac{\big(\Lambda_{\lambda\lambda}-\bar\Lambda_{\lambda\lambda}+\Sigma_{\lambda\lambda}-\bar\Sigma_{\lambda\lambda}\big)}{\big(\Lambda_{\lambda\lambda}+\bar\Lambda_{\lambda\lambda}+\Sigma_{\lambda\lambda}+\bar\Sigma_{\lambda\lambda}\big)}
\quad \text{and} \quad Y_{\lambda\lambda}^r=\frac{Y_{\lambda\lambda}}{\big(\Lambda_{\lambda\lambda}+\bar\Lambda_{\lambda\lambda}+\Sigma_{\lambda\lambda}+\bar\Sigma_{\lambda\lambda}\big)}~,
\end{split}
\end{equation}
where,  $(Y\equiv \rho,\bar\rho,\Sigma,\bar\Sigma)$ with $\lambda\in\{0,\parallel,\perp\}$\,. Knowing $\Phi_i^{meas}$ form the above equations, the \T and \CPT violating parameters in mixing can be obtained from the following equations:
\begin{equation}
\label{eps}
\begin{split}
&\epsilon_1=-\Big(\frac{2}{\sin 2\varphi_\lambda^{meas}}\Big)\Big[\frac{\bar\rho^r_{\lambda\lambda}}{1-C_\lambda}-\frac{\rho^r_{\lambda\lambda}}{1+C_\lambda}-\frac{\Sigma^r_{\lambda\lambda}-\bar\Sigma^r_{\lambda\lambda}}{\sqrt{1-C_\lambda^2}}\,\sin\Phi_\lambda^{meas}\Big]\,,\\
&\epsilon_2=-(\rho^r_{\lambda\lambda}+\bar\rho^r_{\lambda\lambda})+\sqrt{1-C_\lambda^2}\, \sin\Phi_\lambda^{meas}\,,\\
&\epsilon_3=-C_\lambda-\csc\Phi_\lambda^{meas}\;\Big[\rho^r_{\lambda\lambda}\,\sqrt{\frac{1-C_\lambda}{1+C_\lambda}}-\bar\rho^r_{\lambda\lambda}\,\sqrt{\frac{1+C_\lambda}{1-C_\lambda}}\Big]\,.
\end{split}
\end{equation}
Thus we solve for twelve of the fourteen unknown parameters. To obtain the solutions we have inverted the expressions for $\Lambda'_{\lambda\lambda}$, $\Sigma'_{\lambda\lambda}$ and $\rho'_{\lambda\lambda}$ in Eqs. \eqref{prime_to_mode} and \eqref{prime_to_conj} and then use the definitions of those dummy-observables from Eq. \eqref{dummy obs}. It should be noticed that each of $\epsilon_j$ ($j\in\{1,2,3\}$) can obtained in three ways since $\lambda\in\{0,\parallel,\perp\}$. This fact will be used later to find some relations among observables. Now, the remaining two angular quantities $\omega_{\perp i}$ ($i\in\{0,\parallel\}$) are calculated in the following way:
\begin{equation}
\label{omega_pi}
\begin{split}
&\Lambda_{\perp i}+\Sigma_{\perp i}=\Lambda'_{\perp i}+\Sigma'_{\perp i}=-2 \text{ Im } (A_\perp A_i^*) \quad\text{(Using Eqs. \eqref{dummy obs} and \eqref{prime_to_mode})} \\
&\implies \omega_{\perp i}=\text{Arg}[A_\perp]-\text{Arg}[A_i]= \sin^{-1}\Bigg(-\frac{\Lambda_{\perp i}+\Sigma_{\perp i}}{2\sqrt{(\Lambda_{\perp \perp}+\Sigma_{\perp \perp})(\Lambda_{i i}+\Sigma_{i i})}}\Bigg)\,.
\end{split}
\end{equation}

\section{Relations among observables:}
\label{sec:ObservableRelations}


In this section we are going to derive complete set of relations among observables for the scenario with \T and \CPT violation in mixing along with \CP violation in both mixing and decay. As discussed before, we have forty eight observables combining mode and conjugate mode, but the number of unknown theoretical parameters are fourteen. Therefore, we must have forty eight minus fourteen equals to thirty four relations among observables. If we simply substitute the solutions of unknown parameters into Eqs. \eqref{dummy obs} -- \eqref{prime_to_conj}, we would overcount the number of independent relations among observables.

Firstly, it is evident from Eq.~\eqref{eps} that each of the $\epsilon_{j}$ ($j\in\{1,2,3\}$) can express in three ways depending on the value of $\lambda$. Hence, one will give the solution for $\epsilon_{j}$ while the rest two can be recast as relations among observables and this happens for each $\epsilon_j$. Thus we have two times three equals to six relations among observables which are the following:
	\begin{equation}
	\label{eps1}
	\begin{split}
	\bigg(\frac{\sin 2\varphi_i^{meas}}{\sin 2\varphi_\perp^{meas}}\bigg)\Big[\frac{\bar\rho^r_{\perp\perp}}{1-C_\perp}&-\frac{\rho^r_{\perp\perp}}{1+C_\perp}-\frac{\Sigma^r_{\perp\perp}-\bar\Sigma^r_{\perp\perp}}{\sqrt{1-C_\perp^2}}\,\sin\Phi_\perp^{meas}\Big]\\
	&=\Big[\frac{\bar\rho^r_{ii}}{1-C_i}-\frac{\rho^r_{ii}}{1+C_i}-\frac{\Sigma^r_{ii}-\bar\Sigma^r_{ii}}{\sqrt{1-C_i^2}}\,\sin\Phi_i^{meas}\Big]
\end{split}
\end{equation}
	\begin{equation}
	\label{eps2}
	\sqrt{1-C_i^2}\, \sin\Phi_i^{meas}-\sqrt{1-C_\perp^2}\, \sin\Phi_\perp^{meas}=(\rho^r_{ii}+\bar\rho^r_{ii})-(\rho^r_{\perp\perp}+\bar\rho^r_{\perp\perp})
	\end{equation}
	\begin{equation}
	\label{eps3}
	\begin{split}
	 (C_i-C_\perp)+\csc\Phi_i^{meas}&\;\Big[\rho^r_{ii}\,\sqrt{\frac{1-C_i}{1+C_i}}-\bar\rho^r_{ii}\,\sqrt{\frac{1+C_i}{1-C_i}}\Big]\\
	&=\csc\Phi_\perp^{meas}\;\Big[\rho^r_{\perp\perp}\,\sqrt{\frac{1-C_\perp}{1+C_\perp}}-\bar\rho^r_{\perp\perp}\,\sqrt{\frac{1+C_\perp}{1-C_\perp}}\Big]
	\end{split}
	\end{equation}
	where, $i\in\{0,\parallel\}$. The above six relations can be interpreted from a different perspective too. Looking at Eqs. \eqref{modA} -- \eqref{eps}, it can be realized that eighteen observables (three for each of $\Lambda_{\lambda\lambda}$, $\Sigma_{\lambda\lambda}$, $\rho_{\lambda\lambda}$, $\bar\Lambda_{\lambda\lambda}$, $\bar\Sigma_{\lambda\lambda}$ and $\bar\rho_{\lambda\lambda}$) have been used to solve for twelve different quantities (three for each $|A_\lambda|$, $|\bar A_\lambda|$, $\epsilon_j$ and $\sin\varphi_\lambda^{meas}$). Hence, eliminating the unknown quantities, one should get eighteen minus twelve equals to six relations among observables which are given by Eqs. \eqref{eps1} -- \eqref{eps3}.
	
	  Secondly, As mentioned in the last section, we have used only five angular quantities so far (three of $\Phi_\lambda$ and two of $\omega_{\perp i}$). The rest four ($\omega_{\parallel 0}$ and three of $\bar \omega_{\lambda\sigma}$), as defined in Eq. \eqref{angular}, will now be used to find four relations among observables. Let us first express these angles in terms of observables as follows:	
	\begin{equation}
	\label{omega_pa0}
	\begin{split}
	\Lambda_{\parallel0}+&\Sigma_{\parallel 0}=\Lambda'_{\parallel 0}+\Sigma'_{\parallel 0}=2 \text{ Re } (A_\parallel A_0^*) \quad\text{(Using Eqs. \eqref{dummy obs} and \eqref{prime_to_mode})} \\
	&\implies \omega_{\parallel 0}=\text{Arg}[A_\parallel]-\text{Arg}[A_0]= \cos^{-1}\Bigg(\frac{\Lambda_{\parallel 0}+\Sigma_{\parallel 0}}{2\sqrt{(\Lambda_{\parallel \parallel}+\Sigma_{\parallel \parallel})(\Lambda_{0 0}+\Sigma_{0 0})}}\Bigg)\,,
	\end{split}
	\end{equation}
	
		\begin{equation}
		\label{omegabar_pa0}
		\begin{split}
		\bar\Lambda_{\parallel0}+&\bar\Sigma_{\parallel 0}=\Lambda'_{\parallel 0}-\Sigma'_{\parallel 0}=2 \text{ Re } (\bar A_\parallel \bar A_0^*) \quad\text{(Using Eqs. \eqref{dummy obs} and \eqref{prime_to_mode})}\\
		&\implies \bar\omega_{\parallel 0}=\text{Arg}[\bar A_\parallel]-\text{Arg}[\bar A_0]= \cos^{-1}\Bigg(\frac{\bar\Lambda_{\parallel 0}+\bar\Sigma_{\parallel 0}}{2\sqrt{(\bar\Lambda_{\parallel \parallel}+\bar\Sigma_{\parallel \parallel})(\bar\Lambda_{0 0}+\bar\Sigma_{0 0})}}\Bigg)\,,
		\end{split}
		\end{equation}
		
	\begin{equation}
	\label{omegabar_pi}
	\begin{split}
	\bar\Lambda_{\perp i}+&\bar\Sigma_{\perp i}=\Lambda'_{\perp i}-\Sigma'_{\perp i}=2 \text{ Im } (\bar A_\perp \bar A_i^*)\quad\text{(Using Eqs. \eqref{dummy obs} and \eqref{prime_to_mode})}\\
	&\implies \bar\omega_{\perp i}=\text{Arg}[\bar A_\perp]-\text{Arg}[\bar A_i]= \sin^{-1}\Bigg(\frac{\bar\Lambda_{\perp i}+\bar\Sigma_{\perp i}}{2\sqrt{(\bar\Lambda_{\perp \perp}+\bar\Sigma_{\perp \perp})(\bar\Lambda_{i i}+\bar\Sigma_{i i})}}\Bigg)\,,
	\end{split}
	\end{equation}
	where, $i\in\{0,\parallel\}$. Now, from the definitions of these angles, as shown in Eq. \eqref{angular},  it easy to establish the following relations:
		\begin{equation}
		\label{angl_rel}
	\bar\omega_{\lambda\sigma}=\varphi_\lambda^{meas}-\varphi_\sigma^{meas}+\omega_{\lambda\sigma}	\qquad \text{and} \qquad
	\omega_{\parallel 0}=\omega_{\perp 0}-\omega_{\perp \parallel}	
		\end{equation}
    where, 	$(\lambda,\sigma)\in\{0,\parallel,\perp\}$ and hence the first equation contains three relation. Using the expressions of $\omega_{\parallel 0}$ and $\bar \omega_{\lambda\sigma}$ from Eqs. \eqref{omega_pa0} -- \eqref{omegabar_pi} and finding the expressions for $\varphi_\lambda^{meas}$ from Eq. \eqref{sinphi} one can get four relations among observables from Eq. \eqref{angl_rel}.

     Thirdly, using $\Lambda'_{\lambda\sigma}$ and $\Sigma'_{\lambda\sigma}$ for ($\lambda\neq\sigma$), we have solved for two independent angular quantities $\omega_{\perp i}$, as given by Eq. \eqref{omega_pi}, and established four independent relations among observables, given by Eq. \eqref{angl_rel}. However, we see from Eqs. \eqref{prime_to_mode} and \eqref{prime_to_conj} that there are total twelve equations involving  $\Lambda'_{\lambda\sigma}$ and $\Sigma'_{\lambda\sigma}$ with ($\lambda\neq\sigma$) combining the mode and conjugate mode. Therefore, we must have six more relations involving them. Actually, so far we have used $(\Lambda'_{\lambda\sigma}+\Sigma'_{\lambda\sigma})$ for mode and $(\Lambda'_{\lambda\sigma}-\Sigma'_{\lambda\sigma})$ for conjugate mode separately. Now, equating the expressions for $(\Lambda'_{\lambda\sigma}+\Sigma'_{\lambda\sigma})$ from both mode and conjugate mode and repeating it for $(\Lambda'_{\lambda\sigma}-\Sigma'_{\lambda\sigma})$ too we derive the rest six relations as:
     \begin{equation}
     \label{Lam_rel}
     \begin{split}
     &\Lambda_{\lambda\sigma}+\Sigma_{\lambda\sigma}=(1-2\epsilon_3)\;(\bar\Lambda_{\lambda\sigma}-\bar\Sigma_{\lambda\sigma})-2\epsilon_1\;\bar\eta_{\lambda\sigma}+2\epsilon_2\;\bar\rho_{\lambda\sigma}\,,\\
     &\bar\Lambda_{\lambda\sigma}+\bar\Sigma_{\lambda\sigma}=(1+2\epsilon_3)\;(\Lambda_{\lambda\sigma}-\Sigma_{\lambda\sigma})+2\epsilon_1\;\eta_{\lambda\sigma}+2\epsilon_2\;\rho_{\lambda\sigma},
     \end{split}
      \end{equation}
     where, $(\lambda,\sigma)\in\{0,\parallel,\perp\}$ and $\sigma\neq\lambda$\,. Here, one has to use the solutions for $\epsilon_j$ from Eq. \eqref{eps}. It should be noticed that each of two equations in Eq. \eqref{Lam_rel} contains three relations for three different combination of $\lambda$ and $\sigma$ with $\sigma\neq\lambda$\,.

    Fourthly, all the expressions related to $\rho'_{\lambda\lambda}$ have already been used in solving the unknown parameters and deducing the first six relations among observables. But none of the expressions involving $\rho'_{\lambda\sigma}$ with $\lambda\neq\sigma$ has been used yet. Let us first rewrite $\rho'_{\lambda\sigma}$ in terms of observables and the measured angles ($\varphi_\lambda^{meas}$ and $\omega_{\lambda\sigma}$) as follows:
  \begin{equation}
  \label{rhopi}
  \begin{split}
    \rho'_{\perp i}&= -\text{ Re }\Big[\xi(A_\perp^* \bar A_i+A_i^* \bar A_\perp)\Big] \\
    &= -\sqrt{(\Lambda_{\perp \perp}+\Sigma_{\perp \perp})(\bar\Lambda_{i i}+\bar\Sigma_{i i})}\;\cos\;(\varphi^{meas}_i-\omega_{\perp i})\\
    &\qquad\qquad\qquad\qquad-\sqrt{(\Lambda_{ii}+\Sigma_{ii})(\bar\Lambda_{\perp \perp}+\bar\Sigma_{\perp \perp})}\;\cos\;(\varphi^{meas}_\perp+\omega_{\perp i})\,,\quad\,
    \end{split}
    \end{equation}
    \begin{equation}
    \label{rhopa0}
    \begin{split}
    \rho'_{\parallel 0}&= \text{ Im }\Big[\xi(A_\parallel^* \bar A_0+A_0^* \bar A_\parallel)\Big] \\
    &=\sqrt{(\Lambda_{\parallel \parallel}+\Sigma_{\parallel \parallel})(\bar\Lambda_{0 0}+\bar\Sigma_{0 0})}\;\sin\; (\varphi^{meas}_0-\omega_{\parallel 0})\\
    &\qquad\qquad\qquad\qquad+\sqrt{(\Lambda_{0 0}+\Sigma_{0 0})(\bar\Lambda_{\parallel \parallel}+\bar\Sigma_{\parallel \parallel})}\;\sin\; (\varphi^{meas}_\parallel+\omega_{\parallel 0})\,,
    \end{split}
    \end{equation}
     where, $i\in\{0,\parallel\}$. Now using the six equations involving $\rho'_{\lambda\sigma}$ with $\lambda\neq\sigma$ from Eqs. \eqref{prime_to_mode} and \eqref{prime_to_conj} six more  relations among observables can be obtained with the help of Eq. \eqref{eps}.
     
     Lastly, the expressions for $\eta'_{\lambda\sigma}\,\,\forall\,\,(\lambda,\sigma)\in\{0,\parallel,\perp\}$ have not been utilized so far since $\eta_{\lambda\sigma}$ and $\bar{\eta}_{\lambda\sigma}$ become non-measurable in the systems with vanishing $\Delta\Gamma$ (like $B_d^0$). That is why we have tried to eliminate them from most of our solutions and relations (although Eq. \eqref{Lam_rel} contains them). Nonetheless, one can overcome the problem for systems with vanishing $\Delta\Gamma$ as well as find the rest of the relations in case of general $P^0-\bar P^0$ systems. We have to express $\eta'_{\lambda\sigma}\,\,\forall\,\,(\lambda,\sigma)\in\{0,\parallel,\perp\}$ in terms of observables and the measured angles ($\varphi_\lambda^{meas}$ and $\omega_{\lambda\sigma}$) first (like $\rho'$ in the last paragraph) as follows:
    \begin{align}
    \label{etaii}
     &\eta'_{ii}=\text{ Re }\big(\xi A_i^* \bar A_i\big)
     = \sqrt{(\Lambda_{ii}+\Sigma_{ii})(\bar\Lambda_{i i}+\bar\Sigma_{i i})}\;\cos\varphi_i^{meas}\,,\\
     \label{etapp}
      &\eta'_{\perp\perp}=-\text{ Re }\big(\xi A_i^* \bar A_i\big)
      = -\sqrt{(\Lambda_{\perp\perp}+\Sigma_{\perp\perp})(\bar\Lambda_{\perp \perp}+\bar\Sigma_{\perp \perp})}\;\cos\varphi_\perp^{meas}\,,\\
      \label{etapi}
      &\eta'_{\perp i}= \text{ Im }\Big[\xi(A_\perp^* \bar A_i+A_i^* \bar A_\perp)\Big] \nonumber\\
      &\;\,\quad= \sqrt{(\Lambda_{\perp \perp}+\Sigma_{\perp \perp})(\bar\Lambda_{i i}+\bar\Sigma_{i i})}\;\sin\;(\varphi^{meas}_i-\omega_{\perp i})\nonumber\\
      &\qquad\qquad\qquad\qquad +\sqrt{(\Lambda_{ii}+\Sigma_{ii})(\bar\Lambda_{\perp \perp}+\bar\Sigma_{\perp \perp})}\;\sin\;(\varphi^{meas}_\perp+\omega_{\perp i})\,,\\
      \label{etapa0}
      &\eta'_{\parallel 0}= \text{ Re }\Big[\xi(A_\parallel^* \bar A_0+A_0^* \bar A_\parallel)\Big] \nonumber\\
      &\;\quad=\sqrt{(\Lambda_{\parallel \parallel}+\Sigma_{\parallel \parallel})(\bar\Lambda_{0 0}+\bar\Sigma_{0 0})}\;\cos\; (\varphi^{meas}_0-\omega_{\parallel 0})\nonumber\\
      &\qquad\qquad\qquad\qquad+\sqrt{(\Lambda_{0 0}+\Sigma_{0 0})(\bar\Lambda_{\parallel \parallel}+\bar\Sigma_{\parallel \parallel})}\;\cos\; (\varphi^{meas}_\parallel+\omega_{\parallel 0})\,,
       \end{align}
     where $i\in\{0,\parallel\}$. Now, substituting the above relations into to twelve equations involving $\eta'_{\lambda\sigma}\,\,\forall\,\,(\lambda,\sigma)\in\{0,\parallel,\perp\}$ in Eqs. \eqref{prime_to_mode} and \eqref{prime_to_conj}, the remaining twelve relations among observables can be established. For vanishing $\Delta\Gamma$, those twelve relations can be used for theoretical estimation of $\eta_{\lambda\sigma}$ and $\bar\eta_{\lambda\sigma}$ which can used in Eq. \eqref{Lam_rel} to verify those observable relations. Thus excluding the last twelve relations, we have total twenty two observable relations in vanishing $\Delta\Gamma$ scenario, whereas in general cases we have total thirty four relations among observables. Some of these relations will get violated only if there exists direct violation of \CPT (i.e. violation in the decay itself.)

\section{Special cases:}
\label{sec:Splcs}

In this section, we will study following three special cases using our formalism.

\subsection{SM scenario:}
\label{SM}
In SM scenario, there is no violation of \T  (apart from \CP violating effects) and  \CPT in mixing. Hence, $\epsilon_j=0\,\forall\,j\in\{1,2,3\}$ which readily infer from Eqs. \eqref{prime_to_mode} and \eqref{prime_to_conj} that
\begin{equation}
\label{mode_conj_rel_SM}
\Lambda_{\lambda\sigma}=\bar\Lambda_{\lambda\sigma}\,,\quad
\eta_{\lambda\sigma}=\bar\eta_{\lambda\sigma}\,,\quad
\Sigma_{\lambda\sigma}=-\bar\Sigma_{\lambda\sigma}\,,\quad
\rho_{\lambda\sigma}=\bar \rho_{\lambda\sigma}\,,
\end{equation}
where $(\lambda,\sigma)\in\{0,\parallel,\perp\}$. It should be kept in mind that the forty eight equations in Eqs. \eqref{prime_to_mode} and \eqref{prime_to_conj} have been recast as solutions of fourteen theoretical parameter, as given in Sec. \ref{sec:Solution}, and thirty four relations among observables, as described in Sec. \ref{sec:ObservableRelations}. Therefore, the twenty four relations in Eq. \eqref{mode_conj_rel_SM} are also embedded in the solutions or relations among observables. But it would take a bit more algebraic complexity to dig them out from there and so we simply derive them from Eqs. \eqref{prime_to_mode} and \eqref{prime_to_conj}.

The other constrain in SM is that each of helicity amplitudes for mode and the conjugate mode is equal to each other (i.e. $A_\lambda=\bar A_\lambda$). Equating the modulus of helicity amplitude one gets the following three relations from Eqs. \eqref{modA} and \eqref{mode_conj_rel_SM}:
\begin{align}
\label{Sig_SM}
&\Sigma_{\lambda\lambda}=0 \qquad\qquad\forall\; \lambda\in\{0,\parallel,\perp\}\,,\\
\label{modA_SM}
\text{and hence, } 
&|A_\lambda|=\sqrt{\Lambda_{\lambda\lambda}}\qquad\qquad\forall\; \lambda\in\{0,\parallel,\perp\}\,.
\end{align}
On the other hand, equating the phases one would get the following three relations from Eq. \eqref{angular}:
\begin{equation}
\label{sig_lam_SM}
\omega_{\lambda\sigma}=\bar\omega_{\lambda\sigma}\;\;\forall\; (\lambda,\sigma)\in\{0,\parallel,\perp\}\;\;\implies \Sigma_{\parallel 0}=0\;\;\text{and}\;\;\Lambda_{\perp i}=0 \text{ with }i\in\{0,\parallel\}.
\end{equation}
Here, we have used the Eq. \eqref{omega_pi} and Eqs. \eqref{omega_pa0} -- \eqref{omegabar_pi} for the expressions of $\omega_{\lambda\sigma}$ and $\bar\omega_{\lambda\sigma}$. The expressions for $\omega_{\perp i}$ and $\omega_{\parallel 0}$ in this scenario turn out to be following which will be used later:
\begin{equation}
\label{omega_pi_SM}
\omega_{\perp i}=\sin^{-1}\Big(-\frac{\Sigma_{\perp i}}{2\sqrt{\Lambda_{\perp\perp}\Lambda_{ii}}}\Big) \text{  and  }\omega_{\parallel 0}=\cos^{-1}\Big(\frac{\Lambda_{\parallel 0}}{2\sqrt{\Lambda_{\parallel\parallel}\Lambda_{00}}}\Big)\,.
\end{equation}
From Eq. \eqref{angular}, we also get that $\varphi_\lambda^{meas}=-2\beta$. Combining this information with Eqs. \eqref{sinphi}, \eqref{XandC}, \eqref{mode_conj_rel_SM} and \eqref{Sig_SM} results in following two relations:
\begin{equation}
\label{rho_lam_SM}
\frac{\rho_{ii}}{\Lambda_{ii}}=-\frac{\rho_{\perp\perp}}{\Lambda_{\perp\perp}}\quad\text{ with }i\in\{0,\parallel\},
\end{equation}
along with the expression of $\sin2\beta$ as: 
\begin{equation}
\label{sin2beta_SM}
\sin2\beta=-\sin\varphi_\lambda^{meas}=-\Big(\frac{\rho_{00}}{\Lambda_{00}}\Big)\,.
\end{equation}
Thus first part of Eq. \eqref{angl_rel} (i.e. $\bar\omega_{\lambda\sigma}=\varphi_\lambda^{meas}-\varphi_\sigma^{meas}+\omega_{\lambda\sigma}$) gets satisfied automatically. After a couple of discussions we will come back to the second part of the equation.

Now, substituting the Eqs. \eqref{rhopi} and \eqref{rhopa0} into Eq. \eqref{prime_to_mode} and using the expressions of angular quantities $\omega_{\perp i}$, $\omega_{\parallel 0}$ and $\varphi_\lambda^{meas}$ from Eqs. \eqref{omega_pi_SM} and \eqref{sin2beta_SM} along with Eqs. \eqref{mode_conj_rel_SM} and \eqref{Sig_SM}, one arrive at the following three relations:
\begin{align}
\label{rho_lam_SM_1}
\frac{\rho_{\perp i}^2}{4\Lambda_{\perp\perp}\Lambda_{ii}-\Sigma_{\perp i}^2}&=\frac{\Lambda_{00}^2-\rho_{00}^2}{\Lambda_{00}^2} \quad\text{ with }i\in\{0,\parallel\}\,,\\
\label{rho_lam_SM_2}
\text{and, }\qquad\frac{\rho_{\parallel 0}}{\Lambda_{\parallel 0}}&=\frac{\rho_{00}}{\Lambda_{00}}\,.
\end{align}

In the same way, using the expressions for $\eta'_{\lambda\sigma}$ in Eqs. \eqref{etaii} -- \eqref{etapa0}, the following six relations for $i\in\{0,\parallel\}$ can be achieved with the help of a bit of algebraic and trigonometric operations:
\begin{eqnarray}
\label{etabylam_SM}
&\displaystyle\frac{\eta_{ii}^{}}{\Lambda_{ii}}= \frac{\eta_{\parallel 0}^{}}{\Lambda_{\parallel 0}}=-\frac{\eta_{\perp\perp}^{}}{\Lambda_{\perp\perp}}\,,\\
\label{etabyrho_SM}
&\displaystyle\frac{\eta_{\perp i}^{}}{\rho_{\perp i}^{}}+\frac{\eta_{\parallel 0}^{}}{\rho_{\parallel 0}^{}}=0\,,\\
\label{lamsq_SM}
&\displaystyle\eta_{\parallel 0}^2+\rho_{\parallel 0}^2=\Lambda_{\parallel 0}^2\, ,
\end{eqnarray}

Finally, we use the last part of Eq. \eqref{angl_rel} (i.e. $\omega_{\parallel 0}=\omega_{\perp 0}-\omega_{\perp \parallel}$) to reach the last relation:
\begin{equation}
\label{lampa0_SM}
\Lambda_{\parallel 0}=\frac{1}{2\Lambda_{\perp\perp}}\Big[\Sigma_{\perp 0}\Sigma_{\perp\parallel}+\rho_{\perp 0}^{}\rho_{\perp\parallel}^{}\Big(\frac{\Lambda_{00}^2}{\Lambda_{00}^2-\rho_{00}^2}\Big)\Big]
\end{equation}

Thus we have six unknown parameters (three of $|A_\lambda|$, two of $\omega_{\perp i}$ and one $\beta$) in this case to solve for which are given by Eqs. \eqref{modA_SM}, \eqref{omega_pi_SM} (first part) and \eqref{sin2beta_SM}. Therefore, one should get a complete set of forty two independent relations among observables which consists of twenty four in Eq. \eqref{mode_conj_rel_SM}, three in each of Eqs. \eqref{Sig_SM}, \eqref{sig_lam_SM} and \eqref{etabylam_SM}, two in each of Eqs. \eqref{rho_lam_SM}, \eqref{rho_lam_SM_1}, \eqref{etabyrho_SM} and one in each of Eqs. \eqref{rho_lam_SM_2}, \eqref{lamsq_SM} and \eqref{lampa0_SM} respectively. All the other expressions in Sec. \ref{sec:Solution} and \ref{sec:ObservableRelations} satisfy automatically. Except the twenty four relations in Eq. \eqref{mode_conj_rel_SM}, the other eighteen relation have already been discussed in Ref. \cite{karan2}. These relations will get violated by the presence of direct \CP violation or some \CPT non-conserving new Physics effects. 

\subsection{SM plus direct CP violation:}
\label{SM+CP}
In this case also $\epsilon_j=0\,\forall\,j\in\{1,2,3\}$ which immediately imply from Eqs. \eqref{prime_to_mode} and \eqref{prime_to_conj} that
\begin{equation}
\label{mode_conj_rel_CP}
\Lambda_{\lambda\sigma}=\bar\Lambda_{\lambda\sigma}\,,\quad
\eta_{\lambda\sigma}=\bar\eta_{\lambda\sigma}\,,\quad
\Sigma_{\lambda\sigma}=-\bar\Sigma_{\lambda\sigma}\,,\quad
\rho_{\lambda\sigma}=\bar \rho_{\lambda\sigma}\,,
\end{equation}
where $(\lambda,\sigma)\in\{0,\parallel,\perp\}$ like the SM scenario. However, the presence of direct \CP violation in this case makes the helicity amplitudes of the mode to be different from that of the conjugate mode. Thus, here we have eleven theoretical parameters (three for each of $|A_\lambda|$, $|\bar A_\lambda|$ and $\varphi_\lambda^{meas}$ respectively and two of $\omega_{\perp i}$) which follows from Sec. \ref{sec:Solution} and Eq. \eqref{mode_conj_rel_CP} as:
\begin{equation}
\label{solv_CP}
\begin{split}
&|A_\lambda|=\sqrt{\Lambda_{\lambda\lambda}+\Sigma_{\lambda\lambda}}\,,\quad |\bar A_\lambda|=\sqrt{\Lambda_{\lambda\lambda}-\Sigma_{\lambda\lambda}}\,,\quad \sin\Phi_\lambda^{meas}= \frac{\rho_{\lambda\lambda}}{\sqrt{\Lambda_{\lambda\lambda}^2-\Sigma_{\lambda\lambda}^2}}\,,\\
&\qquad\qquad\qquad\omega_{\perp i}= \sin^{-1}\Bigg(-\frac{\Lambda_{\perp i}+\Sigma_{\perp i}}{2\sqrt{(\Lambda_{\perp \perp}+\Sigma_{\perp \perp})(\Lambda_{i i}+\Sigma_{i i})}}\Bigg)\,.
\end{split}
\end{equation}
It should be noticed that the other two solutions for $\sin\Phi_\lambda^{meas}$ from the cubic equation in Eq. \eqref{sinphi} are imaginary in general since $\Lambda_{\lambda\lambda}>\Sigma_{\lambda\lambda}$ is required for positive definiteness of $|\bar A_\lambda|$. With the help of Eqs. \eqref{mode_conj_rel_CP} and \eqref{solv_CP}, the expressions for $\epsilon_j$ in Eq. \eqref{eps} vanish automatically.

In this case, number of independent relations among observables is forty eight minus eleven equals to thirty seven. Among them twenty four are listed in Eq. \eqref{mode_conj_rel_CP}. The remaining thirteen can be found in the following way: 1) four relations can be found from Eq. \eqref{angl_rel}, 2) three can be established by substituting Eqs. \eqref{rhopi} and \eqref{rhopa0} in to Eq. \eqref{prime_to_mode}, 3) the last six can be obtained by replacing Eqs. \eqref{etaii} -- \eqref{etapa0} into Eq. \eqref{prime_to_mode}. After a bit of mathematical jugglery, these thirteen relations can be described in the following form:
\begin{equation}
\label{cpv1}
\hspace*{-20mm}\Lambda_{\lambda\lambda}=\sqrt{\Sigma_{\lambda\lambda}^2+\rho_{\lambda\lambda}^2+\eta_{\lambda\lambda}^2}\,,
\end{equation}
\begin{equation}
\label{cpv2}
\hspace*{-10mm}\Bigg[\frac{(\frac{\rho_{\sigma\sigma}}{\Lambda_{\sigma\sigma}+\Sigma_{\sigma\sigma}})+(\frac{\rho_{\lambda\lambda}}{\Lambda_{\lambda\lambda}+\Sigma_{\lambda\lambda}})-2(\frac{\rho_{\lambda\sigma}}{\Lambda_{\lambda \sigma}+\Sigma_{\lambda\sigma}})}{(\frac{\eta_{\sigma\sigma}}{\Lambda_{\sigma\sigma}+\Sigma_{\sigma\sigma}})+(\frac{\eta_{\lambda\lambda}}{\Lambda_{\lambda\lambda}+\Sigma_{\lambda\lambda}})-2(\frac{\eta_{\lambda\sigma}}{\Lambda_{\lambda\sigma}+\Sigma_{\lambda\sigma}})}\Bigg]=-\Bigg[\frac{\big(\frac{\eta_{\lambda\lambda}}{\Lambda_{\lambda\lambda}+\Sigma_{\lambda\lambda}}\big)-\big(\frac{\eta_{\sigma\sigma}}{\Lambda_{\sigma\sigma}+\Sigma_{\sigma\sigma}}\big)}{\big(\frac{\rho_{\lambda\lambda}}{\Lambda_{\lambda\lambda}+\Sigma_{\lambda\lambda}}\big)-\big(\frac{\rho_{\sigma\sigma}}{\Lambda_{\sigma\sigma}+\Sigma_{\sigma\sigma}}\big)}\Bigg]\,,
\end{equation}
\begin{align}
\label{cpv3}
\hspace*{-1mm}4\Bigg[\frac{(\Lambda_{\lambda\lambda}+\Sigma_{\lambda\lambda})(\Lambda_{\sigma\sigma}+\Sigma_{\sigma\sigma})}{(\Lambda_{\lambda\sigma}+\Sigma_{\lambda\sigma})^2}\Bigg]-\Bigg[\frac{\big(\frac{\rho_{\lambda\lambda}}{\Lambda_{\lambda\lambda}+\Sigma_{\lambda\lambda}}\big)+\big(\frac{\rho_{\sigma\sigma}}{\Lambda_{\sigma\sigma}+\Sigma_{\sigma\sigma}}\big)-2\big(\frac{\rho_{\lambda\sigma}}{\Lambda_{\lambda\sigma}+\Sigma_{\lambda\sigma}}\big)}{\big(\frac{\eta_{\lambda\lambda}}{\Lambda_{\lambda\lambda}+\Sigma_{\lambda\lambda}}\big)-\big(\frac{\eta_{\sigma\sigma}}{\Lambda_{\sigma\sigma}+\Sigma_{\sigma\sigma}}\big)}\Bigg]^2=1\,,
\end{align}
\begin{align}
\label{cpv4}
\Bigg[ 4\bigg(\frac{\eta_{\lambda\lambda}\rho_{\sigma\sigma}-\eta_{\sigma\sigma}\rho_{\lambda\lambda}}{\Lambda_{\lambda\sigma}^2-\Sigma_{\lambda\sigma}^2}\bigg)&+\bigg\{\frac{\big(\frac{\rho_{\lambda\lambda}}{\Lambda_{\lambda\lambda}+\Sigma_{\lambda\lambda}}\big)+\big(\frac{\rho_{\sigma\sigma}}{\Lambda_{\sigma\sigma}+\Sigma_{\sigma\sigma}}\big)-2\big(\frac{\rho_{\lambda\sigma}}{\Lambda_{\lambda\sigma}+\Sigma_{\lambda\sigma}}\big)}{\big(\frac{\eta_{\lambda\lambda}}{\Lambda_{\lambda\lambda}+\Sigma_{\lambda\lambda}}\big)-\big(\frac{\eta_{\sigma\sigma}}{\Lambda_{\sigma\sigma}+\Sigma_{\sigma\sigma}}\big)}\bigg\}\Bigg]^2\nonumber\\
&\hspace*{3cm}=4\Bigg[\frac{(\Lambda_{\lambda\lambda}-\Sigma_{\lambda\lambda})(\Lambda_{\sigma\sigma}-\Sigma_{\sigma\sigma})}{(\Lambda_{\lambda\sigma}-\Sigma_{\lambda\sigma})^2}\Bigg]-1\,,
\end{align}
\begin{align}
\label{cpv5}
&\Bigg[\frac{(\Lambda_{\parallel 0}+\Sigma_{\parallel 0})^2}{(\Lambda_{0 0}+\Sigma_{0 0})(\Lambda_{\parallel \parallel}+\Sigma_{\parallel \parallel})}\Bigg]-\Bigg[\frac{(\Lambda_{\perp 0}+\Sigma_{\perp 0})(\Lambda_{\perp \parallel}+\Sigma_{\perp \parallel})(\Lambda_{\parallel 0}+\Sigma_{\parallel 0})}{(\Lambda_{\perp \perp}+\Sigma_{\perp \perp})(\Lambda_{0 0}+\Sigma_{0 0})(\Lambda_{\parallel \parallel}+\Sigma_{\parallel \parallel})}\Bigg]\nonumber\\
&\hspace*{15mm}=4-\Bigg[\frac{(\Lambda_{\perp 0}+\Sigma_{\perp 0})^2}{(\Lambda_{\perp \perp}+\Sigma_{\perp \perp})(\Lambda_{0 0}+\Sigma_{0 0})}\Bigg]-\Bigg[\frac{(\Lambda_{\perp \parallel}+\Sigma_{\perp \parallel})^2}{(\Lambda_{\perp \perp}+\Sigma_{\perp \perp})(\Lambda_{\parallel \parallel}+\Sigma_{\parallel \parallel})}\Bigg]\,,
\end{align}
where, $(\lambda,\sigma)\in\{0,\parallel,\perp\}$ and $\lambda\neq\sigma$. It should be noticed that each of the four equations from Eq. \eqref{cpv1} to Eq. \eqref{cpv4} contains three relations for different values of $\lambda$ and $\sigma$ with $\lambda$ not being equal to $\sigma$, and Eq. \eqref{cpv5} contains only one. Violation of these relations would definitely imply existence of \CPT violating new Physics phenomenon.

\subsection{SM plus T and CPT violation in mixing:}
\label{SM+CPT}

In this case, one can follow the entire procedure described in Sec. \ref{sec:Solution} and \ref{sec:ObservableRelations} to get the solutions of theoretical parameters and find the relations among observables. But to reach the expressions in the form of Ref. \cite{karan2}, that already discusses this scenario, one has to  encounter various algebraic complexities.

As there is no direct \CP violation in this case, the helicity amplitudes for mode and conjugate mode will be equal to each other (like SM). All of the three $\varphi_\lambda^{meas}$ become $-2\beta$ too. Therefore, there will be total nine unknown parameters (three of $|A_\lambda|$, two of $\omega_{\perp i}$, one $\beta$ and three of $\epsilon_j$). It implies that the total number of independent relations among observables is forty eight minus nine equal to thirty nine. It should be noticed from Sec. \ref{SM} that $\Sigma_{\lambda\lambda}$, $\Sigma_{\parallel 0}$, $\Lambda_{\perp i}$, $(\frac{\rho_{ii}}{\Lambda_{ii}}+\frac{\rho_{\perp\perp}}{\Lambda_{\perp\perp}})$, $(\frac{\eta_{ii}}{\Lambda_{ii}}+\frac{\eta_{\perp\perp}}{\Lambda_{\perp\perp}})$, $(\frac{\rho_{ii}}{\eta_{ii}}-\frac{\rho_{\perp\perp}}{\eta_{\perp\perp}})$, etc. were zero in SM case. However, inverting Eq. \eqref{prime_to_mode} one can find that they are $\mathcal O(\epsilon_j)$ in the present scenario. Since we are keeping track up to the linear order terms in $\epsilon_j$, any quadratic term involving the above expressions will be neglected. The same rule applies for the observables of conjugate mode too.  

At first, using the equality of helicity amplitudes for mode and conjugate mode, one achieve the following six relations from Eqs. \eqref{modA}, \eqref{omega_pi}, \eqref{omega_pa0} -- \eqref{angl_rel} (first part):
\begin{equation}
\label{CPT1}
\Lambda_{\lambda\lambda}+\Sigma_{\lambda\lambda}=\bar \Lambda_{\lambda\lambda}+\bar \Sigma_{\lambda\lambda},\quad \Lambda_{\perp i}+\Sigma_{\perp i}=-(\bar \Lambda_{\perp i}+\bar \Sigma_{\perp i}), \quad \Lambda_{\parallel 0}+\Sigma_{\parallel 0}=\bar \Lambda_{\parallel 0}+\bar \Sigma_{\parallel 0},
\end{equation} 
where, $\lambda\in\{0,\parallel,\perp\}$ and $i\in\{0,\parallel\}$. Along with the above six relations we also get the expressions for five unknown parameters as:
\begin{equation}
\label{modA_CPT}
|A_\lambda|=\sqrt{\Lambda_{\lambda\lambda}+\Sigma_{\lambda\lambda}}\quad\text{and}\quad
\omega_{\perp i}= \sin^{-1}\Bigg(-\frac{\Lambda_{\perp i}+\Sigma_{\perp i}}{2\sqrt{(\Lambda_{\perp \perp}+\Sigma_{\perp \perp})(\Lambda_{i i}+\Sigma_{i i})}}\Bigg)\,.
\end{equation}

At second, we use the three expressions for $\varphi_\lambda^{meas}\;(=-2\beta)$ from Eq. \eqref{sinphi} and six relations in Eqs. \eqref{eps1} -- \eqref{eps3} that equate the different expressions for $\epsilon_j$ ($j\in\{1,2,3\}$). These nine expressions can be paraphrased as one equation for $\sin2\beta$ and eight relations among observables as follows:
\begin{equation}
\label{sin2beta_CPT}
\sin2\beta=
-\frac{1}{2}\Big(\frac{\rho_{00}}{\Lambda_{00}}-\frac{\rho_{\perp\perp}}{\Lambda_{\perp\perp}}\Big)\,,
\end{equation}

\begin{equation}
\label{CPT2}
\Big(\frac{\rho_{00}}{\Lambda_{00}}-\frac{\rho_{\perp\perp}}{\Lambda_{\perp\perp}}\Big)=\Big(\frac{\bar\rho_{00}}{\bar\Lambda_{00}}-\frac{\bar\rho_{\perp\perp}}{\bar\Lambda_{\perp\perp}}\Big)\,,
\end{equation}

\begin{equation}
\label{CPT3}
\frac{\Sigma_{ 00}}{\Lambda_{ 00}}=\frac{\Sigma_{\parallel\parallel}}{\Lambda_{\parallel\parallel}}\,,\qquad \frac{\rho_{ 00}}{\Lambda_{ 00}}=\frac{\rho_{\parallel\parallel}}{\Lambda_{\parallel\parallel}}\,, \qquad \frac{\bar\Sigma_{ 00}}{\bar\Lambda_{ 00}}=\frac{\bar\Sigma_{\parallel\parallel}}{\bar\Lambda_{\parallel\parallel}}\,,\qquad \frac{\bar\rho_{ 00}}{\bar\Lambda_{ 00}}=\frac{\bar\rho_{\parallel\parallel}}{\bar\Lambda_{\parallel\parallel}}\,,
\end{equation}

\begin{equation}
\label{rho00pp_CPT}
\hspace*{10mm}\Big(\frac{\rho_{00}-\bar\rho_{00}}{\Lambda_{00}+\Sigma_{00}}\Big)+\Big(\frac{\rho_{\perp\perp}-\bar\rho_{\perp\perp}}{\Lambda_{\perp\perp}+\Sigma_{\perp\perp}}\Big)=0\,,\quad
\Big(\frac{\Sigma_{00}+\bar\Sigma_{00}}{\Lambda_{00}+\Sigma_{00}}\Big)+\Big(\frac{\Sigma_{\perp\perp}+\bar\Sigma_{\perp\perp}}{\Lambda_{\perp\perp}+\Sigma_{\perp\perp}}\Big)=0\,,
\end{equation}

\begin{equation}
\label{rhosigma00pp_2_CPT}
\Big(\frac{\Sigma_{00}+\bar\Sigma_{00}}{\Lambda_{ 00}+\Sigma_{ 00}}\Big)-\Big(\frac{\rho_{ 00}^{}}{\Lambda_{ 00}+\Sigma_{ 00}}+\frac{\rho_{\perp\perp}^{}}{\Lambda_{\perp\perp}+\Sigma_{\perp\perp}}\Big) \sin 2\beta=0
\end{equation}

\noindent With the help of the above expressions the values  of $\epsilon_2$ and $\epsilon_3$ in Eq. \eqref{eps} can be written as:
\begin{equation}
\label{eps23_CPT}
\epsilon_2=-\frac{1}{2}\Big(\frac{\rho_{ 00}^{}}{\Lambda_{ 00}+\Sigma_{ 00}}+\frac{\rho_{\perp\perp}^{}}{\Lambda_{\perp\perp}+\Sigma_{\perp\perp}}\Big), \qquad
\epsilon_3=\frac{1}{2}\Big(\frac{\Sigma_{ 00}^{}}{\Lambda_{ 00}+\Sigma_{ 00}}+\frac{\Sigma_{\perp\perp}^{}}{\Lambda_{\perp\perp}+\Sigma_{\perp\perp}}\Big).
\end{equation}

Thirdly, substituting the expressions for $\eta'_{\lambda\lambda}$ for $\lambda\in\{0,\parallel,\perp\}$ from Eqs. \eqref{etaii} and \eqref{etapp} into Eqs. \eqref{prime_to_mode} and \eqref{prime_to_conj} one would end up with following six relations:

\begin{equation}
\label{CPT_eta1}
\frac{\eta_{ 00}}{\Lambda_{ 00}}=\frac{\eta_{\parallel\parallel}}{\Lambda_{\parallel\parallel}}\,, \quad \frac{\bar\eta_{ 00}}{\bar\Lambda_{ 00}}=\frac{\bar\eta_{\parallel\parallel}}{\bar\Lambda_{\parallel\parallel}}\,,
\end{equation}

\begin{align}
\label{CPT_eta2}
\Big(\frac{\rho_{00}}{\Lambda_{00}}-\frac{\rho_{\perp\perp}}{\Lambda_{\perp\perp}}\Big)^2+
\Big(\frac{\eta_{00}^{}}{\Lambda_{00}}-\frac{\eta^{}_{\perp\perp}}{\Lambda_{\perp\perp}}\Big)^2=4
\end{align}
\begin{align}
\label{CPT_eta3} \Big(\frac{\eta_{00}^{}}{\Lambda_{00}}-\frac{\eta^{}_{\perp\perp}}{\Lambda_{\perp\perp}}\Big)=\Big(\frac{\bar\eta_{00}^{}}{\bar\Lambda_{00}}-\frac{\bar\eta^{}_{\perp\perp}}{\bar\Lambda_{\perp\perp}}\Big)
\end{align}

\begin{align}
\label{CPT_eta4}
\frac{\rho_{00}^2+\eta_{00}^2}{\Lambda_{00}^2}=\frac{\rho_{ \perp\perp}^2+\eta_{ \perp\perp}^2}{\Lambda_{ \perp\perp}^2}, \quad \frac{\bar\rho_{00}^2+\bar\eta_{00}^2}{\bar\Lambda_{00}^2}=\frac{\bar\rho_{ \perp\perp}^2+\bar\eta_{ \perp\perp}^2}{\bar\Lambda_{ \perp\perp}^2}
\end{align}

\noindent It should be noticed that some of the above relations could be further simplified as: $\Lambda_{\lambda\lambda}=\sqrt{\eta_{\lambda\lambda}^2+\rho_{\lambda\lambda}^2}$ and $\bar\Lambda_{\lambda\lambda}=\sqrt{\bar\eta_{\lambda\lambda}^2+\bar\rho_{\lambda\lambda}^2}$. Nonetheless, to reproduce the relations in Ref. \cite{karan2}, we stick to the former ones only. Using the above relations, the expression for $\epsilon_1$ from Eq. \eqref{eps} can be interpreted as following:
\begin{equation}
\label{eps1_CPT}
\epsilon_1 = -\frac{1}{2}\Big(\frac{\eta_{ ii}^{}}{\Lambda_{ ii}+\Sigma_{ ii}}+\frac{\eta_{\perp\perp}^{}}{\Lambda_{\perp\perp}+\Sigma_{\perp\perp}}\Big)\,.
\end{equation}
With the help of above relations involving $\eta_{\lambda\lambda}$, one can also abandon Eq. \eqref{rhosigma00pp_2_CPT} and recast it as:
\begin{equation}
\label{etaii_CPT}
\Big(\frac{\eta_{ ii}^{}-\bar\eta_{ ii}^{}}{\Lambda_{ ii}+\Sigma_{ ii}}\Big)+\Big(\frac{\eta_{\perp\perp}^{}+\bar\eta_{\perp\perp}^{}}{\Lambda_{\perp\perp}+\Sigma_{\perp\perp}}\Big)=0
\end{equation}

Fourthly, we use the Eq. \eqref{Lam_rel} to arrive at the expressions for $\cos\omega_{\lambda\sigma}$ and $\cos\bar\omega_{\lambda\sigma}$ that can be rewritten as the following six relations:
\small
\begin{align}
\label{SinsqCossqDelta}
\Big[\frac{(\Lambda_{\perp i}+\Sigma_{\perp i})^2}{(\Lambda_{ii}+\Sigma_{ii})(\Lambda_{\perp\perp}+\Sigma_{\perp\perp})}\Big]+
4 X_i^2\, \Lambda_{ ii}^2\,\Lambda_{\perp\perp}^2\Big[\frac{(\Lambda_{ii}+\Sigma_{ii})(\Lambda_{\perp\perp}+\Sigma_{\perp\perp})}{(\Lambda_{\perp\perp}\Sigma_{ ii}+\Lambda_{ii}\Sigma_{\perp\perp}+2\Lambda_{\perp\perp}\Lambda_{ii})^2}\Big] =4\,,
\end{align}
\begin{align}
\label{SinsqCossqDeltabar}
\Big[\frac{(\bar\Lambda_{\perp i}+\bar\Sigma_{\perp i})^2}{(\bar\Lambda_{ii}+\bar\Sigma_{ii})(\bar\Lambda_{\perp\perp}+\bar\Sigma_{\perp\perp})}\Big]+
4 \bar X_i^2\, \bar\Lambda_{ ii}^2\,\bar\Lambda_{\perp\perp}^2\Big[\frac{(\bar\Lambda_{ii}+\bar\Sigma_{ii})(\bar\Lambda_{\perp\perp}+\bar\Sigma_{\perp\perp})}{(\bar\Lambda_{\perp\perp}\bar\Sigma_{ ii}+\bar\Lambda_{ii}\bar\Sigma_{\perp\perp}+2\bar\Lambda_{\perp\perp}\bar\Lambda_{ii})^2}\Big] =4\,,
\end{align}
\normalsize
\begin{equation}
\label{sig_lam_CPT}
\frac{\Sigma_{\parallel 0}}{\Lambda_{\parallel 0}}=\frac{\Sigma_{0 0}}{\Lambda_{0 0}}\quad \text{and} \quad \frac{\bar\Sigma_{\parallel 0}}{\bar\Lambda_{\parallel 0}}=\frac{\bar\Sigma_{0 0}}{\bar\Lambda_{0 0}}\,,
\end{equation}

 \begin{align*}
\hspace*{-0.3cm}\text{where, }{
	X_i=\Big[\frac{\big(\Lambda_{\perp i}-\Sigma_{\perp i}\big)\big(\Lambda_{\perp\perp}\Sigma_{ ii}+\Lambda_{ ii}\Sigma_{\perp\perp}\big)+2\big(\Lambda_{ ii}\Lambda_{\perp\perp}\Lambda_{\perp i}-\Sigma_{ ii}\Sigma_{\perp\perp}\Sigma_{\perp i}\big)}{\big(\eta_{\perp\perp}^{} \rho_{ii}^{}-\eta_{ii}^{} \rho_{\perp\perp}^{}\big)\big(\Lambda_{ii}+\Sigma_{ii}\big)\big(\Lambda_{\perp\perp}+\Sigma_{\perp\perp}\big)}\Big]}, 
\end{align*}
\begin{align*}
\hspace*{-0.3cm}\text{and, }
\oln{X}_i=\Big[\frac{\big(\oln{\Lambda}_{\perp i}-\oln{\Sigma}_{\perp i}\big)\big(\oln{\Lambda}_{\perp\perp}\oln{\Sigma}_{ ii}+\oln{\Lambda}_{ ii}\oln{\Sigma}_{\perp\perp}\big)+2\big(\oln{\Lambda}_{ ii}\oln{\Lambda}_{\perp\perp}\oln{\Lambda}_{\perp i}-\oln{\Sigma}_{ ii}\oln{\Sigma}_{\perp\perp}\oln{\Sigma}_{\perp i}\big)}{\big(\oln{\eta}_{\perp\perp}^{} \oln{\rho}_{ii}^{}-\oln{\eta}_{ii}^{} \oln{\rho}_{\perp\perp}^{}\big)\big(\oln{\Lambda}_{ii}+\oln{\Sigma}_{ii}\big)\big(\oln{\Lambda}_{\perp\perp}+\oln{\Sigma}_{\perp\perp}\big)}\Big],
\end{align*}
with $i\in\{0,\parallel\}$. It is very important to note that to get a correct $X_i$ or $\bar X_i$ up to $\mathcal O (\epsilon_j)$ one should keep the quadratic terms of $\epsilon_j$ in the numerator and denominator separately while defining $X_i$ or $\bar X_i$, since the leading order terms in both of the numerator and denominator are $\mathcal O (\epsilon_j)$. Notwithstanding, the relations in Eqs. \eqref{SinsqCossqDelta} and \eqref{SinsqCossqDeltabar} can be untangle a bit by using $\cos\omega_{\perp i}=\frac{1}{2}X_i\sqrt{\Lambda_{\perp\perp}\Lambda_{ii}}$ and  $\cos\bar\omega_{\perp i}=\frac{1}{2}\bar X_i\sqrt{\bar\Lambda_{\perp\perp}\bar\Lambda_{ii}}$\,.

Fifthly, by replacing the expressions for $\rho'_{\lambda\sigma}$ with $\lambda\neq\sigma$ from Eqs. \eqref{rhopi} and \eqref{rhopa0} into Eqs. \eqref{prime_to_mode} and \eqref{prime_to_conj}, as described in Sec. \ref{sec:ObservableRelations}, one would end up with following six relations:
\small
\begin{align}
\label{rhopi_CPT}
&\rho_{\perp i}=\frac{1}{2}\Bigg[\frac{\Sigma_{\perp i}}{\Lambda_{ii}\Lambda_{\perp\perp}}\Big\{\rho_{\perp\perp}^{}\big(\Lambda_{ii}+\Sigma_{ii}\big)+\rho_{ii}^{}\big(\Lambda_{\perp\perp}+\Sigma_{\perp\perp}\big)\Big\}-X_i\Big\{\Lambda_{\perp\perp}\eta_{ii}^{}-\Lambda_{ii}\eta_{\perp\perp}^{}\Big\}\Bigg],\\
&\bar\rho_{\perp i}=\frac{1}{2}\Bigg[\frac{\bar\Sigma_{\perp i}}{\bar\Lambda_{ii}\bar\Lambda_{\perp\perp}}\Big\{\bar\rho_{\perp\perp}^{}\big(\bar\Lambda_{ii}+\bar\Sigma_{ii}\big)+\bar\rho_{ii}^{}\big(\bar\Lambda_{\perp\perp}+\bar\Sigma_{\perp\perp}\big)\Big\}-\bar X_i\Big\{\bar\Lambda_{\perp\perp}\bar\eta_{ii}^{}-\bar\Lambda_{ii}\bar\eta_{\perp\perp}^{}\Big\}\Bigg],
\end{align}
\normalsize
\begin{equation}
\label{rho_lam_CPT}
\frac{\rho_{\parallel 0}}{\Lambda_{\parallel 0}}=\frac{\rho_{0 0}}{\Lambda_{0 0}}\quad \text{and} \quad \frac{\bar\rho_{\parallel 0}}{\bar\Lambda_{\parallel 0}}=\frac{\bar\rho_{0 0}}{\bar\Lambda_{0 0}}\,,
\end{equation}
where,  $i\in\{0,\parallel\}$. Similarly, substituting the expressions for $\eta'_{\lambda\sigma}$ with $\lambda\neq\sigma$ from Eqs. \eqref{etapi} and \eqref{etapa0} into Eqs. \eqref{prime_to_mode} and \eqref{prime_to_conj}, would lead to the following six relations:
\small
\begin{align}
\label{etapi_CPT}
&\eta_{\perp i}=\frac{1}{2}\Bigg[\frac{\Sigma_{\perp i}}{\Lambda_{ii}\Lambda_{\perp\perp}}\Big\{\eta_{\perp\perp}^{}\big(\Lambda_{ii}+\Sigma_{ii}\big)+\eta_{ii}^{}\big(\Lambda_{\perp\perp}+\Sigma_{\perp\perp}\big)\Big\}+X_i\Big\{\Lambda_{\perp\perp}\rho_{ii}^{}-\Lambda_{ii}\rho_{\perp\perp}^{}\Big\}\Bigg], \\
&\bar\eta_{\perp i}=\frac{1}{2}\Bigg[\frac{\bar\Sigma_{\perp i}}{\bar\Lambda_{ii}\bar\Lambda_{\perp\perp}}\Big\{\bar\eta_{\perp\perp}^{}\big(\bar\Lambda_{ii}+\bar\Sigma_{ii}\big)+\bar\eta_{ii}^{}\big(\bar\Lambda_{\perp\perp}+\bar\Sigma_{\perp\perp}\big)\Big\}+\bar X_i\Big\{\bar\Lambda_{\perp\perp}\bar\rho_{ii}^{}-\bar\Lambda_{ii}\bar\rho_{\perp\perp}^{}\Big\}\Bigg], 
\end{align}
\normalsize
\begin{equation}
\label{eta_lam_CPT}
\frac{\eta_{\parallel 0}}{\Lambda_{\parallel 0}}=\frac{\eta_{0 0}}{\Lambda_{0 0}}\quad \text{and} \quad \frac{\bar\eta_{\parallel 0}}{\bar\Lambda_{\parallel 0}}=\frac{\bar\eta_{0 0}}{\bar\Lambda_{0 0}}\,.
\end{equation}

Finally, second part of Eq. \eqref{angl_rel} $(\omega_{\parallel 0}=\omega_{\perp 0}-\omega_{\perp \parallel})$ indicates the last independent relation as:
\begin{align}
\label{Delta0minuspara}
\Big(\Lambda_{0\parallel}+\Sigma_{0\parallel}&\Big)-\frac{1}{2}\Big[\frac{(\Lambda_{\perp 0}+\Sigma_{\perp 0})(\Lambda_{\perp \parallel}+\Sigma_{\perp \parallel})}{(\Lambda_{\perp\perp}+\Sigma_{\perp\perp})}\Big]\nonumber \\
&=\Big[\frac{2X_0 X_\parallel\, \Lambda_{ 00}\Lambda_{\parallel\parallel}\,\Lambda_{\perp\perp}^2(\Lambda_{00}+\Sigma_{00})(\Lambda_{\parallel\parallel}+\Sigma_{\parallel\parallel})(\Lambda_{\perp\perp}+\Sigma_{\perp\perp})}{(\Lambda_{\perp\perp}\Sigma_{00}+\Lambda_{00}\Sigma_{\perp\perp}+2\Lambda_{\perp\perp}\Lambda_{00})(\Lambda_{\perp\perp}\Sigma_{\parallel\parallel}+\Lambda_{\parallel\parallel}\Sigma_{\perp\perp}+2\Lambda_{\perp\perp}\Lambda_{\parallel\parallel})}\Big],
\end{align}

Thus, the expressions for nine theoretical parameters (three of $|A_\lambda|$, two of $\omega_{\perp i}$, one $\beta$ and three of $\epsilon_j$) in this scenario are given by five equations in Eq. \eqref{modA_CPT}, two expressions in Eq. \eqref{eps23_CPT} and one in each of Eq. \eqref{sin2beta_CPT} and \eqref{eps1_CPT}. On the other hand, the thirty nine relation among observables are presented as: a) six relations in Eq. \eqref{CPT1}, b) seven equations from Eq. \eqref{CPT2} -- \eqref{rho00pp_CPT} (we have not counted Eq. \eqref{rhosigma00pp_2_CPT} since it has been recast as Eq. \eqref{etaii_CPT}), c) six expressions from Eq. \eqref{CPT_eta1} -- \eqref{CPT_eta4} and  d) twenty relations from Eq. \eqref{etaii_CPT} -- \eqref{Delta0minuspara}. If some of these relations do not hold true, that will indicate the presence of \CPT or \CP violation in decay itself.

\section{Phenomenology:}
\label{sec:Phenomenology}
 Various experiments have been performed so far in order to probe \CPT violation in neutral meson mixing. Although these experiments measure tiny non-zero values for \CPT violating parameters, they become consistent to zero within $2\sigma$ due to the presence of experimental error bars with comparable size. In case of kaon system, \CPT asymmetry is measured from the semileptonic ($\pi^+l^-\bar\nu_l$, $\pi^-l^+\nu_l$) decay modes of $K^0$ and $\bar K^0$ to estimate the \CPT-violating complex parameter $\delta$ whose real and imaginary parts are directly proportional to $\epsilon_1$ and $\epsilon_2$ respectively in our notation. From the data of KTeV collaboration \cite{ktev}, the real and imaginary parts of this parameter are estimated to be: $\Re(\delta)=(2.51\pm2.25)\times10^{-4}$ and $\Im(\delta)=(-1.5\pm1.6)\times10^{-5}$ which agree with \CPT conservation. In case of $D^0-\bar D^0$ system, \CPT asymmetry, which is constructed by comparing the time dependent decay probabilities of the modes $D^0\to K^-\pi^+$ and $\bar D^0\to K^+\pi^-$, has been measured by FOCUS collaboration \cite{focus}. This measurement leads to the estimation of \CPT-violating complex parameter $\xi$, whose real and imaginary parts are proportional to $\epsilon_1$ and $\epsilon_2$ respectively, to be: $\Re (\xi)\, y-\Im (\xi)\, x=0.0083\pm0.0065\pm0.0041$ where, $x=\frac{\Delta M}{\Gamma}$ and $y=\frac{\Delta \Gamma}{2\Gamma}$. The first measurement of \CPT violation in $B^0-\bar B^0$ system was performed by BaBar collaboration \cite{CPTVBmix_BaBar1,CPTVBmix_BaBar2}. The last update on it has been carried out by Belle collaboration \cite{expts1}. For this purpose, they have fitted the time dependent decay rate of the chain: $\Upsilon(4S)\to B^0_d\,\bar B^0_d\to f_{rec}\, f_{tag}$ , where one of the $B$-meson decays to reconstructed final state $f_{rec}$ at time $t_{rec}$ and the other one decays at time $t_{tag}$ to a final state $f_{tag}$, that distinguishes between $B^0$ and $\bar B^0$. Several hadronic and semileptonic decay modes of $B^0_d$ ($J/\psi K_S$, $J/\psi K_L$, $D^-\pi^+$, $D^{*-}\pi^+$, $D^{*-}\rho^+$ and $D^{*-}l^+\nu_l$) have been used in this case to find the experimental value for the \CPT-violating complex parameter $z$ as: $\Re (z)=(1.9\pm3.7\pm3.3)\times10^{-2}$ and $\Im(z)=(-5.7\pm3.3\pm3.3)\times10^{-3}$ which are consistent with zero. Similarly, using the time dependent decay rate to a \CP eigenstate for the mode $B^0_s\to J/\psi K^+ K^-$, the \CPT violation in $B_s^0$ system has been measured by LHCb collaboration \cite{expts2} as: $\Re (z)=-0.022\pm0.033\pm0.005$ and $\Im(z)=0.004\pm0.011\pm0.002$ that also agrees with conservation of \CPT. The dependence of the parameter $z$ on $\epsilon_1$ and $\epsilon_2$ is already given by Eq. \eqref{epsdefs}. On the other hand, direct and indirect \CP violations as well as \T violation in neutral meson mixing have also been measured at different occasions \cite{pdg}. However, during all these measurements, \CPT has always been assumed to be preserved which makes \T and \CP violations equivalent to each other.

Due to unavailability of enough phase space, the kaon system cannot decay to any vector meson. Nevertheless, for $D^0$, $B^0_d$ and $B_s^0$ systems, several decay modes with two vectors in final state, to which both the pseudoscalar meson and its antiparticle can decay, are accessible. As this paper addresses direct \CP violation with the decays $P^0\to f$ and $\bar P^0\to f$, we should focus on the final states $f$ which are \CP eigenstates. The modes that can be considered for $D^0$ system are $K^*\bar K^*$,  $\rho^0\phi$, $K^{*+} K^{*-}$, $\rho^0\rho^0$, $\rho^+\rho^-$, etc. Similarly, for $B^0_d$ systems, modes like $\phi\phi$, $\rho^0\rho^0$, $\rho^+\rho^-$, $\omega\omega$, $K^*\bar K^*$, $K^{*+} K^{*-}$, $D_s^{*+}D_s^{*-}$, etc. can be used.  For $B_s^0$ system also, the modes like $\phi\phi$, $\rho^0\rho^0$, $\rho^+\rho^-$, $K^*\bar K^*$, $K^{*-} K^{*-}$, $D_s^{*+}D_s^{*-}$, $J/\psi\, \phi$, etc. should be used.  Many of these modes have already been studied by several experimental groups \cite{pdg}. But time dependence of the decay rates for these modes has not been used so far in connection with probing \CPT violation.

Having said that, let us now consider the final states consisting of two vectors with no definite \CP. The general analysis, described up to sec. \ref{sec:ObservableRelations}, does not require the final state $f$ to be a \CP eigenstate. So, one can use it for any final state $f$ to which both $P^0$ and $\bar P^0$ can decay. In that case, the decay rates for the channels $P^0\to f$ and $\bar P^0\to f$ could be very different from each other and the SM behaviour would exactly be described by the scenario depicted in sec. \ref{SM+CP}. But the equations of sec. \ref{SM} and \ref{SM+CPT} must not be used for this particular scenario since these two sections inevitably presume the final state $f$ to be a \CP eigenstate while imposing the condition $A_\lambda=\bar A_\lambda$ for \CP being conserved. Generally, if $f$ does not represent a \CP eigenstate, only one of the channels between $P^0\to f$ and $\bar P^0 \to f$ becomes dominant due to additional CKM suppression or presence of extra $W$-boson propagator in the leading order Feynman diagram of the other process. For example, this happens in the decay channels like $\bar K^{*}\rho^0$ or $ K^{*\pm}\rho^\mp$ for $D^0$ system, $D^{*\pm}K^{*\mp}$, $D^{*\pm}\rho^{\mp}$, $\phi K^*$, $J/\psi K^*$, etc., modes in case of $B^0_d$ systems and $D_s^{*\pm}\rho^{\mp}$, $\phi K^*$, $J/\psi K^*$, etc., channels for $B^0_s$ systems. Nevertheless, it does not mean that all the observables for the disfavoured modes will be small enough in magnitude. To elaborate, let us consider a SM scenario (i.e. $\epsilon_j=0$) with $|\bar A_\lambda/A_\lambda|\ll 1$. Then, Eqs. \eqref{dummy obs} - \eqref{prime_to_conj} readily imply that $(\rho_{\lambda\sigma}/\Lambda_{\lambda\sigma})\approx (\bar\rho_{\lambda\sigma}/\bar\Lambda_{\lambda\sigma})\ll 1$, $(\eta_{\lambda\sigma}/\Lambda_{\lambda\sigma})\approx (\bar\eta_{\lambda\sigma}/\bar\Lambda_{\lambda\sigma})\ll 1$ and $\Lambda_{\lambda\sigma}\approx\bar\Lambda_{\lambda\sigma}\approx\Sigma_{\lambda\sigma}\approx-\bar\Sigma_{\lambda\sigma}$. So, if we neglect the terms $\eta_{\lambda\sigma}$ and $\rho_{\lambda\sigma}$ with respect to $\Lambda_{\lambda\sigma}$ and $\Sigma_{\lambda\sigma}$ and then integrate the time dependent decay rates (given by Eqs. \eqref{VVDecayRate} and \eqref{ConjVVDecay}) over time, we find the branching fractions for $P^0\to f$ and $\bar P^0\to f$ to be: 
$\displaystyle \text{Br}\,(P^0\to f)\approx\Big(\frac{1}{\Gamma^2-\Delta \Gamma^2/4}+\frac{1}{\Gamma^2+\Delta M^2}\Big)\sum_{\lambda\leq\sigma}\Lambda_{\lambda\sigma}g_\lambda g_\sigma$ and 
$\displaystyle \text{Br}\,(\bar P^0\to f)\approx\Big(\frac{1}{\Gamma^2-\Delta \Gamma^2/4}-\frac{1}{\Gamma^2+\Delta M^2}\Big)\sum_{\lambda\leq\sigma}\Lambda_{\lambda\sigma}g_\lambda g_\sigma\,.$ This clearly shows that  $\Lambda_{\lambda\sigma}$ might not be small in this special case but still branching fraction for the disfavoured transition could be small due to cancellation between two large contributions coming from integration of $e^{-\Gamma t}\cosh(\Delta\Gamma t/2)$ and $e^{-\Gamma t}\cos(\Delta M t)$ terms depending on the values of $\Delta\Gamma$ and $\Delta M$ whereas for the dominant mode these contributions  add up to a larger value. However,  due to presence of large contributions from $\Lambda_{\lambda\sigma}$ and $\Sigma_{\lambda\sigma}$, precise measurement for $\eta_{\lambda\sigma}$ and $\rho_{\lambda\sigma}$ might be problematic in these kind of channels. 

Now, by examining the decay products of the produced vectors, the helicity component for the two vector final state can be obtained. Then studying angular analysis and time dependence of the decay rates for $P^0\to f$ and $\bar P^0\to f$, all the observables can be measured \cite{VV1}. However, due to existence of several relations among observables, the \T, \CP and \CPT violating parameters can be expressed in various ways involving different observables. But in experiment, all these observables can be measured separately and hence several measurements for same theoretical parameter can be performed simultaneously which would help to reduce the error bar allowing a more precise measurement for the unknown parameter. Thus better results on \CPT violation should be expected from future runs of LHCb and Belle II.

\section{Conclusion:}
\label{sec:Conclusion}

In conclusion, we have studied the behaviour of observables for neutral meson decaying to two vectors in the presence of \T, \CP and \CPT violation in mixing as well as \CP violation in decay. Polarizations of final state with two vectors  provide us a large number of observables in these modes. The final state should be chosen in such a way that both $P^0$ and $\bar P^0$ can decay to it. We extract all of the fourteen unknown theoretical parameters in terms of the observables and then discuss the procedure to establish the complete set of independent relations among observables containing thirty four equations. These relations can be used as the smoking gun signal to prove the existence of direct violation of \CPT (if any) since those effects only can lead to non-obedience of them. Additionally, we explore three special cases e.g. SM case, SM plus direct \CP violation scenario and SM plus \T and \CPT violation in mixing case. Using our new formalism, we derive the expressions for unknown theoretical parameters and construct the complete set of independent relations among observables too in each special case. Experimental verification for each of the sets will signify the existence of some particular type of Physics. For example, the set of relations in SM plus \T and \CPT violation in mixing scenario can be applied to probe direct violation of  \CPT or \CP, the set of relations in SM plus \CP case can be implemented to confirm the existence of any \CPT violating new Physics (direct or indirect), whereas the set of observable relations in SM scenario should be used to detect direct \CP violation or \CPT non-conserving new Physics.

\section*{Acknowledgement}
The author thanks Rahul Sinha and Abinash Kumar Nayak for some useful discussions. The author also thanks SERB India,  grant no: CRG/2018/004971, for the financial support.

\end{document}